\documentclass[aps,twocolumn,showpacs,amsmath,amssymb,superscriptaddress]{revtex4}

\usepackage{epsfig}
\usepackage{graphicx}
\usepackage{dcolumn}
\usepackage{bm}
\usepackage{hyperref}

\begin{document}

\title{Metric of a tidally perturbed spinning black hole}

\author{Nicol\'as  Yunes}
\email{yunes@gravity.psu.edu}
\affiliation{Institute for Gravitational Physics and Geometry,
             Center for Gravitational Wave Physics,
             Department of Physics, The Pennsylvania State University,
             University Park, PA 16802-6300}

\author{Jos\'e A. Gonz\'alez}
\email{jose.gonzalez@uni-jena.de}
\affiliation{Theoretical Physics Institute, University of Jena,
  Max-Wien-Platz 1, 07743, Jena, Germany}

\date{$$Id: notes.tex,v 1.5 2005/07/29 13:40:21 yunes Exp $$}

\pacs{
04.25.Dm,   
04.25.Nx,   
04.30.Db,  
95.30.Sf    
}


\preprint{IGPG-05/11-3}

%
\newcommand\be{\begin{equation}}
\newcommand\ba{\begin{eqnarray}}
\newcommand\ee{\end{equation}}
\newcommand\ea{\end{eqnarray}}
\newcommand\p{{\partial}}
\newcommand\remove{{{\bf{THIS FIG. OR EQS. COULD BE REMOVED}}}}
%

\begin{abstract}
  We explicitly construct the metric of a Kerr black hole that is
  tidally perturbed by the external universe in the slow-motion
  approximation. This approximation assumes that the external universe
  changes slowly relative to the rotation rate of the hole, thus
  allowing the parameterization of the Newman-Penrose scalar $\psi_0$
  by time-dependent electric and magnetic tidal tensors. This
  approximation, however, does not constrain how big the spin of the
  background hole can be and, in principle, the perturbed metric can
  model rapidly spinning holes. We first generate a potential by
  acting with a differential operator on $\psi_0$. From this potential
  we arrive at the metric perturbation by use of the Chrzanowski
  procedure in the ingoing radiation gauge. We provide explicit
  analytic formulae for this metric perturbation in Kerr coordinates,
  where the perturbation is finite at the horizon. This perturbation
  is parametrized by the mass and Kerr spin parameter of the
  background hole together with the electric and magnetic tidal
  tensors that describe the time evolution of the perturbation
  produced by the external universe. In order to make the metric
  accurate far away from the hole, these tidal tensors should be
  determined by asymptotically matching this metric to another one
  valid far from the hole. The tidally perturbed metric constructed
  here could be useful in initial data constructions to describe the
  metric near the horizons of a binary system of spinning holes. This
  perturbed metric could also be used to construct waveforms and study
  the absorption of mass and angular momentum by a Kerr black hole
  when external processes generate gravitational radiation.

\end{abstract}

\maketitle

\section{Introduction}
\label{intro}
Gravitational wave observatories, such as LIGO and VIRGO, have the
potential to study black holes in the strong field regime
\cite{Cutler:2002me}. These black holes are expected to be
immersed in a sea of gravitational perturbations that will alter the
gravitational field of the background hole. Even though tidal
perturbations are expected to be small relative to the background,
they will be important in some astrophysical scenarios when attempting
to provide an accurate description of the non-linear dynamical orbital
evolution of bodies around this background. The need for high accuracy
in the description of the orbital evolution derives from the fact that
gravitational wave observatories are extremely sensitive to the phase
of the gravitational waves emitted by the system. Therefore, since
this phase is directly related to the orbital evolution, in some
astrophysical scenarios it is necessary to take these tidal effects
into consideration.

The study of gravitational perturbations around Kerr black holes is
important for several reasons. First, it is of astrophysical interest
to study the flux of mass and angular momentum across a perturbed Kerr
horizon
\cite{Poisson:2004cw,Poisson:1994yf,Tagoshi:1997gq,Alvi:2001mx}, which
can be calculated through manipulations of the tidally perturbed
metric computed in this paper. Although this flux might be small for
equal-mass binaries, in extreme-mass ratio inspirals (EMRI) up to
$5\%$ of the total energy might be absorbed by the background hole.
This absorption might slow down the orbital evolution increasing the
duration of the gravitational wave signal
\cite{Martel:2003jj,Hughes:2001jr}. Space-based detectors, such as
LISA, will be able to observe and measure the gravitational waveforms
of EMRIs, since they have particularly low noise in the low frequency
band where such inspirals are common. Therefore, precise knowledge of
the gravitational waveform including the tidal perturbations effects
might be important in data analysis \cite{Glampedakis:2005cf}.
Finally, the explicit formulae of this paper might be useful to
compute initial data near the horizons of a binary system of spinning
holes. For example, Refs.~\cite{Alvi:1999cw,Yunes:2005nn} make use of
such explicit formulae for the non-spinning case to construct initial
data via asymptotic matching.  This data might be useful to the
numerical relativity community because it derives from an approximate
solution to the Einstein equation and, thus, we expect it to
accurately describe the gravitational field of the system up to
uncontrolled remainders.

In this paper, we analytically construct explicit formulae for the
metric of a tidally perturbed Kerr hole, where the perturbations of
the external universe are assumed to vary slowly in a well-defined
sense.  Metric perturbations for non-spinning holes have been studied
in
Refs.~\cite{Death:1974o,Death:1975,Deathbook,Alvi:1999cw,Poisson:2005pi}
using the Regge-Wheeler formalism \cite{Regge-Wheeler}.  However, this
method is difficult to implement for spinning holes because the metric
will now depend on both radius and angle $\theta$ in a non-trivial
way, rendering the Einstein equations very difficult to solve. For
this reason, we use the Chrzanowski procedure
\cite{1975PhRvD..11.2042C,Ori:2002uv} to construct the metric
perturbation from the Newman-Penrose (NP) scalar $\psi_0$. This
procedure allows us to calculate the metric in the so-called ingoing
radiation gauge (IRG), which is suitable to study gravitational
perturbations near the outer horizon ($r_+$), since there the metric
is transverse and traceless.

We will work in the slow-motion approximation, described in detail in
Refs.~\cite{Burke-Thorne,Burke,Death:1974o,Death:1975,Thorne:1984mz,Deathbook,Poisson:2004cw,Poisson:2005pi},
where we assume that the rate of change of the curvature of the
external universe is small relative to the rotation rate of the
background black hole, which in principle could be extremal. The
external universe is completely arbitrary in that sense, as long as it
respects the slow-motion approximation.  For example, in the case
where the external universe is given by a second black hole in a
quasicircular orbit around the background hole, this approximation is
valid as long as their orbital separation is sufficiently large. In
particular, this separation must be at least greater than the inner
most stable circular orbit
(ISCO)\cite{Baumgarte00a,Grandclement:2001ed}, so that the binary is
still in a quasicircular orbit. In that case, the curvature generated
by the second hole would correspond to the external universe, which
will change slowly as long as the orbital velocity is sufficiently
small.  In this sense, the slow-motion approximation will hold for
astrophysically realistic binaries as long as they are sufficiently
separated.

Assuming this approximation to be valid, Poisson \cite{Poisson:2004cw}
has computed $\psi_0$ in the neighborhood of a spinning hole.  First,
the Weyl tensor of the spacetime is re-expressed in terms of the
electric and magnetic tidal tensors of the external universe. Using
the slow-motion approximation, Poisson argues that these tensors will
be spatially coordinate independent if evaluated at sufficiently large
distances from the worldline of the hole. With this tensor, the
asymptotic form of $\psi_0$, denoted by $\tilde{\psi}_0$, is computed
far from the hole by projecting the Weyl tensor onto the Kinnersly
tetrad.  This scalar will be a combination of slowly varying functions
of time, which will be parametrized via the electric and magnetic
tidal tensors, and scalar functions of the spatial coordinates, which
will be given by the tetrad. The asymptotic form of $\psi_0$ now
allows for the construction of an ansatz for $\psi_0$, which consists
of its asymptotic form $\tilde{\psi}_0$ multiplied by a set of
undetermined function of radius $R_m(r)$. These functions must satisfy
the asymptotic condition $R_m \to 1$ as $r \gg r_+$, as well as the
Teukolsky equation. This last condition is a differential constraint
on $R_m(r)$ which can be solved for explicitly, thus allowing for the
full determination of $\psi_0$. In this manner, the final expression
for $\psi_0$ is obtained and is now valid close to the horizon as
well, in particular, approaching the perturbations generated by the
external universe sufficiently far from the hole's worldline.

Once $\psi_0$ has been calculated, we can apply the Chrzanowski
procedure to compute the metric perturbation, still in the slow-motion
approximation. This calculation contains two parts: the computation of
a potential ($\Psi$) and the determination of the metric perturbation
($h_{ab}$) from $\Psi$. In principle, one might think that it would be
easier to try to compute $h_{ab}$ directly from $\psi_0$. Chrzanowski
\cite{1975PhRvD..11.2042C} attempted this by applying a differential
operator onto $\psi_0$.  However, Wald \cite{Wald:1978vm} discovered
that doing this leads to a physically different gravitational
perturbation from that represented by $\psi_0$. Cohen and Kegeles
\cite{Kegeles:1979an} showed that by constructing a Hertz-like
potential $\Psi$ from $\psi_0$ first and then applying Chrzanowski's
differential operator to $\Psi$ instead leads to the real metric
perturbation. This is the procedure we will follow to construct the
metric perturbation in this paper.

The construction of $\Psi$ requires the action of a fourth order
differential operator on $\psi_0$, where here we follow Ori
\cite{Ori:2002uv}. This potential is simplified by the use of the
slow-motion approximation that allows us to neglect any time
derivatives of the electric and magnetic tidal tensors.  Once $\Psi$
is calculated, we can apply Chrzanowski's differential operator to
this potential \cite{Kegeles:1979an,Campanelli:1998jv}. In this
manner, we compute the metric of a perturbed spinning hole from a
tidal perturbation described by $\psi_0$ in terms of tidal tensors.
These tensors are unknown functions of time that represent the
external universe and which should be determined by asymptotically
matching this metric to another approximation valid far from the holes
\cite{Alvi:1999cw,Yunes:2005nn}.

The metric computed here, however, has a limited applicability given
by the validity of the slow-motion approximation and the Chrzanowski
procedure. The slow-motion approximation implies that we can neglect
all derivatives of the tidal tensors. Furthermore, since we are
working only to first non-vanishing order in this approximation, it
suffices to consider only the quadrupolar perturbation of the metric,
since the monopolar and dipolar perturbations are identically zero.
In perturbation theory, any $l$ mode in the decomposition of the
perturbation is one order larger than the $l+1$ mode. Therefore, any
higher modes or couplings of the quadrupole to other modes will be of
higher order. The Chrzanowski procedure also possesses a limited
region of validity, given by a region sufficiently close to the event
horizon so that the spatial distance from the horizon to the radius of
curvature of the external universe is small. This restriction is
because the Chrzanowski procedure builds the metric as a linear
perturbation of the background and neglects any non-linear
interactions with the external universe. In particular, this
restriction implies that this metric cannot provide valid information
on the dynamics of the entire spacetime.  However, if this metric is
asymptotically matched to another approximation that is valid far from
the background hole, then the combined global metric will describe the
the $3$-manifold accurately.

We verify the validity of our calculations in several ways. First, we
check that $\psi_0$ indeed satisfies the Teukolsky equation. After
computing $\Psi$, we also check that it satisfies the Teukolsky
equation and the differential constraint that relates $\Psi$ to
$\psi_0$.  Finally, we check that the metric perturbation constructed
with this potential satisfies all of the Einstein equations to the
given order.  We further check that this perturbation is indeed
transverse and traceless in the tetrad frame so that it is suitable
for the study of gravitational perturbations near the horizon.
 
This paper is divided as follows. Sec.~\ref{eric} describes the
slow-motion approximation in detail, summarizes some relevant results
from Ref.~\cite{Poisson:2004cw} and establishes some notation.
Sec.~\ref{ori} computes $\Psi$ from $\psi_0$ while Sec.~\ref{lousto}
calculates $h_{ab}$ from $\Psi$.  Finally, Sec.~\ref{conclusions}
presents some conclusions and points toward future work. In the
appendix, we provide an explicit transformation to Kerr-Schild
coordinates that might be more amenable to numerical implementation.

In the remaining of the paper, we use geometrized units ($G=1$, $c=1$)
and the symbol $O(a)$ stands for terms of order $a$, where $a$ is
dimensionless. Latin indices range from $0$ to $3$, where $0$ is the
time coordinate. The Einstein summation convention is assumed all
throughout the paper, where repeated indices are to be summed over
unless otherwise specified. Tetrad notation will be used, where
indices with parenthesis refer to the tetrad and those without
parenthesis to the components of the tensor. The relational symbol
$\sim$ stands for ``asymptotic to'' as defined in \cite{Bender}, while
the symbols $\ll$ and $\gg$ are also to be understood in the
asymptotic sense. In particular, note that if $f(r)$ is valid for $r
\ll b$, then this function is not valid as $r \to b$. In this paper we
have relied heavily on the use of symbolic manipulation software, such
as MAPLE and MATHEMATICA.

\section{The slow-motion approximation and the NP scalar}
\label{eric}
In this section we will describe the slow-motion approximation in more
detail and discuss the construction of $\psi_0$ due to perturbations
of the external universe. Both the slow-motion approximation and
$\psi_0$ have already been explained in detail and computed by Poisson
in Ref.~\cite{Poisson:2004cw}. Therefore, here we follow this
reference and summarize the most relevant results for this paper while
establishing some notation.

Let us begin by discussing the slow-motion approximation.  Consider a
non-spinning black hole of mass $m_1$ immersed in an external
universe, with radius of curvature ${\cal{R}}$. This external universe
could be given by any object that lives in the exterior of the hole's
horizon, such as a scalar field or another black hole. The slow-motion
approximation requires that the external universe's length scales be
much larger than the hole's scales. In other words, for the
non-spinning case we must have $m_1/{\cal{R}} \ll 1$, since these are
the only scales available.

For concreteness, let us assume that the external universe can be
described by another object of mass $M_{ext}$ and that our hole is in
a quasicircular orbit around it. Then, we have
\be
\label{vel-slow-motion}
\frac{m_1}{{\cal{R}}} \sim \frac{m_1}{m_1+M_{ext}} V^2, \qquad V = \sqrt{
  \frac{m_1+ M_{ext}}{b}},
\ee
where $V$ is the orbital velocity and $b$ is the orbital separation.
There are two ways of enforcing $m_1/{\cal{R}} \ll 1$: the small-body
approximation, where we let $m_1/M_{ext} \ll 1$; and the slow-motion
approximation, where $V \ll 1$. However, in a future paper we might
want to asymptotically match the metric perturbation computed in this
paper to a post-Newtonian (PN) expansion \cite{Blanchet:2002av}, which
requires small velocities. For this reason, we will restrict our
attention to the slow-motion approximation, which implies that we can
only investigate systems that are sufficiently separated. The ISCO is
not a well-defined concept for black hole binaries, but it has been
estimated for non-spinning binary numerically
\cite{Baumgarte00a,Grandclement:2001ed} to be given approximately by
$\omega_{ISCO} M/10$, where $M$ and $\omega_{ISCO}$ are the total mass
of the system and its angular velocity at the ISCO respectively. For
spinning binaries the holes can get closer without plunging, where the
value of the ISCO becomes a function of the spin parameter of the
holes.  Regardless of the type of binary system, the slow-motion
approximation will hold as long as we consider systems that are
separated by at least more than their ISCO so that they are still in a
quasicircular orbit.

The above considerations suffice for non-spinning holes because there
is only one length scale associated with it, $m_1$. However, for
spinning ones we must also take into account the time scale associated
with the intrinsic spin of the hole. The slow motion approximation
does not constraint how large the spin of the background hole could
be. However, in the standard theory, it is usually assumed that
isolated holes will obey $a < m_1$, where $a$ is the rotation
parameter of the hole and $m_1$ is its mass. If the previous
inequality did not hold, then the hole would be tidally disrupted by
centrifugal forces. When the background hole is surrounded by an
accretion disk, however, some configuration might lead to a violation
of the previous condition \cite{Marcus}, but we will not consider
those in this paper. Let us then define a dimensionless rotation
parameter $\chi = a/m_1$, which is now restricted to $0<\chi <1$. The
mass of the hole and this rotation parameter now define a new
timescale, related to the rotation rate of the horizon and given by
\be
\tau_H = \frac{1}{\Omega_H} = \frac{2 m_1}{\chi} ( 1 + \sqrt{1 - \chi^2} ),
\ee
where $\Omega_H$ is the angular velocity of the hole's horizon
\cite{Poisson}. The slow-motion approximation then requires this time
scale to be much smaller than the time scale associated with the
change of the radius of curvature of the external universe,
$\tau_{ext}$, {\textit{i.e.}}\ $\tau_H/\tau_{ext} \ll 1$. This
condition then becomes
\be
m_1/{\cal{R}} \ll \chi,
\ee
since $\tau_{ext} \approx {\cal{R}}$.  We take the above relation as
the precise definition of the slow-motion approximation for spinning
black holes. Following the reasoning that lead to
Eq.~(\ref{vel-slow-motion}), we must then have $V^2 \ll \chi$, which,
for a binary system, means that the orbital velocity of the system
cannot exceed the rotation rate of the background hole. Thus, the
slow-motion approximation implies that, to this order and for spinning
holes, we can also only consider systems that are sufficiently
separated and where the black holes have relatively rapid spins. This
reasoning does not imply that the Schwarzschild limit is incompatible
with the slow motion approximation. Actually,
Poisson~\cite{Poisson:2004cw} showed that when computing certain
quantities, such as $\psi_0$, this limit can be recovered if we work
to higher order in the slow-motion approximation. The precise radius
of convergence of the slow-motion approximation to first order remains
unknown, but an approximate measure of how small a separation the
approximation can tolerate will be studied in a later section.

Now that the slow-motion approximation has been explained in detail,
let us proceed with the construction of $\psi_0$, which is defined by
\be
\psi_0 = C_{abcd} l^a m^b l^c m^d,
\ee
where $C_{abcd}$ is the Weyl tensor of the spacetime and $l^a$ and
$m^a$ are the first and third tetrad vectors. This null vector $m^a$
is not to be confused with the $m$ that we will introduce later in
this section to denote the angular mode of the perturbation. Poisson
works with the Kinnersly tetrad in advanced Eddington-Finkelstein (EF)
coordinates, also known as Kerr coordinates, which are well-behaved at
the outer horizon, given by $r_+=m_1 + (m_1^2 - a^2)^{1/2}$.

The calculation of $\psi_0$ is based on making an ansatz guided by its
asymptotic form far from the worldline of the hole but less than the
radius of curvature of the external universe, {\textit{i.e.}}\ $r_+
\ll r \ll {\cal{R}}$. In this region, the Weyl tensor can be
decomposed into electric ${\cal{E}}_{ab}$ and magnetic
${\cal{B}}_{ab}$ tidal fields, which are slowly varying functions of
advanced time $v$ only.  Thus, in this region, the only spatial
coordinate dependence in $\psi_0$ is given by the tetrad, namely
\be
\label{asymp-psi0}
\tilde{\psi}_0 \sim - \sum_m z_m(v) \; _2Y_2^m(\theta,\phi),
\ee
where the tilde is to remind us that this quantity is the asymptotic
form of $\psi_0$ and where $_2Y_2^m(\theta,\phi)$ are spin-weighted
spherical harmonics, given by
\ba
_2Y^0_2(\theta,\phi) &=& -\frac{3}{2} \sin^2{\theta},
\nonumber \\
_2Y^{\pm 1}_2(\theta,\phi) &=& - \sin{\theta}(\cos{\theta} \mp 1)
e^{\pm i \phi},
\nonumber \\
_2Y^{\pm 2}_2(\theta,\phi) &=& \frac{1}{4} ( 1 \mp 2 \cos{\theta} +
\cos^2{\theta}) e^{\pm 2 i \phi}.
\ea
In Eq.~(\ref{asymp-psi0}), the quantities $z_m(v)$ are complex
combinations of the tidal fields given by $z_m(v) = \alpha_m(v) + i
\beta_m(v)$, where
\ba
\alpha_0(v) &=& {\cal{E}}_{11}(v) + {\cal{E}}_{22}(v),
\nonumber \\
\alpha_{\pm 1}(v) &=& {\cal{E}}_{13}(v) \mp i {\cal{E}}_{23}(v),
\nonumber \\
\alpha_{\pm 2}(v) &=& {\cal{E}}_{11}(v) - {\cal{E}}_{22}(v) \mp 2 i
{\cal{E}}_{12}(v),
\nonumber \\
\beta_0(v) &=& {\cal{B}}_{11}(v) + {\cal{B}}_{22}(v),
\nonumber \\
\beta_{\pm 1}(v) &=& {\cal{B}}_{13}(v) \mp i {\cal{B}}_{23}(v),
\nonumber \\
\beta_{\pm 2}(v) &=& {\cal{B}}_{11}(v) - {\cal{B}}_{22}(v) \mp 2 i {\cal{B}}_{12}(v).
\ea
Note that in this region, $\psi_0$ is independent of radial coordinate.

With Eq.~(\ref{asymp-psi0}) at hand, Poisson makes an ansatz for the
functional form of $\psi_0$, namely
\be
\label{ansatz}
\psi_0 = -\sum_m z_m(v) R_m(r) _2Y^m_2(\theta,\phi),
\ee
where $R_m(r)$ is an undetermined function of radius that must satisfy
$R_m(r) \to 1$ as $r \gg r_+$. If Eq.~(\ref{ansatz}) is inserted into
the Teukolsky equation, one obtains a differential equation for
$R_m(r)$. The angular part of the Teukolsky equation is automatically
satisfied by the angular decomposition of $\psi_0$ in spin-weighted
spherical harmonics with eigenvalue $E=4$ (Eq.~$2.10$ in
Ref.~\cite{Teukolsky:1973}). The spatial part of the Teukolsky
equation yields a differential constraint for $R_m(r)$ namely,
\ba
\left\{ x (1 + x) \frac{d^2}{dx^2} \right. && + \left. \left[ 3 ( 2
    x + 1) + 2 i m  \gamma\right] \frac{d}{dx}
\right. 
\nonumber \\
&& \left. + 4 i m \gamma \frac{2 x + 1}{x (1 + x)} \right\} R_m(x) = 0,
\ea
where $x$ is a rescaled version of the radial coordinate given by
\be
x = \frac{ r - r_+}{r_+ - r_-},
\ee
and where the inner and outer horizons are given, respectively, by
$r_{\pm} = m_1^2 \pm ( m_1^2 - a^2)^{1/2}$. We should note that $r_+$
is an event horizon, while $r_-$ is actually an apparent horizon.
Solving this equation~\cite{Poisson:2004cw} one obtains
\be
\label{radial-func}
R_m(r) = A_m x^{-2} (1 + x)^{-2} F(-4,1,-1+2im\gamma;-x),
\ee
where $A_m$ is a normalization constant given by 
\be
A_m =  -\frac{i}{6} m \gamma ( 1 + im\gamma)(1 + 4 m^2 \gamma^2).
\ee
In Eq.~(\ref{radial-func}), the function $F(a,b,c;x)$ is the
hypergeometric function
\be
F(a,b,c;x) = \sum_{n=0}^{\infty} \frac{(a)_n (b)_n}{(c)_n} \frac{x^n}{n!},
\ee
where $(a)_n$ is the Pochammer symbol defined as
\ba
(a)_n &=& a (a+1)(a+2)\ldots(a+n-1) = \frac{(a+n-1)!}{(a-1)!},
\nonumber \\
(a)_0 &=& 1.
\ea
For the present case, the series gets truncated at the fourth power and
we obtain 
\ba
&& F(-4,1,-1+2im\gamma;-x) = \left( 1 + \frac{4}{2 i m \gamma - 1} x 
\right. 
\nonumber \\
&& \left. 
+ \frac{6}{(2 i m \gamma - 1)i m \gamma} x^2 
+ \frac{12}{(2 i m \gamma - 1)i m \gamma(2 i m \gamma +1)}
  x^3 
\right. 
\nonumber \\
&& \left. 
+ \frac{12}{(2 i m  \gamma - 1)i m \gamma(2 i m \gamma +
    1) (2 i m \gamma + 2)} x^4  \right),
\ea
where $\gamma$ is a constant given by 
\be
\gamma = \frac{a}{r_+ - r_-}.
\ee

In this manner, Poisson calculates $\psi_0$, which encodes the
gravitational perturbations of the external universe on a spinning
hole in the slow-motion approximation. The full expression for
$\psi_0$ is then given by
\begin{widetext}
  \ba
\label{psi0}
\psi_0 = -\sum_{m \neq 0} B_m x^{-2} (1 + x)^{-2} && \left( 1 +
    \frac{4}{2 i m \gamma - 1} x + \frac{6}{(2 i m \gamma - 1)i m
      \gamma} x^2 + \frac{12}{(2 i m \gamma - 1)i m \gamma(2 i m
      \gamma +1)} x^3 
\right.
  \nonumber \\
  && \left.  + \frac{12}{(2 i m \gamma - 1)i m \gamma(2 i m \gamma +
      1) (2 i m \gamma + 2)} x^4 \right) \; _2Y^m_2(\theta,\phi), 
\ea
\end{widetext}
where we have used the final abbreviation 
\be
B_m = A_m z_m(v),
\ee
without summing over repeated indices here, to group all terms that
are spatially coordinate independent. We have checked that
Eq.~(\ref{psi0}) indeed satisfies the Teukolsky equation
\cite{Teukolsky:1973} for the $s=2$ mode that corresponds to this
scalar. Note that, in this final expression, summation over the $m=0$
mode is removed because Poisson~\cite{Poisson:2004cw} has shown that it
corresponds to the Schwarzschild limit and, thus, it contributes at a
relative $O(m_1/{\cal{R}})$ higher than all other modes. Also note
that the final expression for $\psi_0$ only contains the quadrupolar
$l=2$ mode, once more because higher multipoles will be smaller by a
relative factor of $m_1/{\cal{R}}$ in the slow-motion approximation.
For this reason, there are no mode couplings in the $\psi_0$ presented
above.

The time-evolution of the tidal perturbation will be exclusively
governed by the electric and magnetic tidal fields. These tensors
should be determined by asymptotically matching the metric
perturbation generated by this $\psi_0$ to another approximation valid
far from the hole. However, in
Refs.~\cite{Thorne:1984mz,Alvi:1999cw,Yunes:2005nn,Poisson:2005pi} it
has been shown that when the external universe is given by another
non-spinning black hole in a quasicircular orbit, these tensors scale
approximately as
\be
\label{approx-z}
z_m \approx \frac{m_2}{b^3}, 
\ee
where $m_2$ is the mass of the other hole (the external universe) and
$b$ is the orbital separation (approximately equal to the radius of
curvature of the external universe). Note that the factor of $b^3$ in
the denominator is necessary to make the tidal tensors dimensionally
correct. Also note that in Eq.~(\ref{approx-z}) we have neglected the
time dependence of the tidal fields, which generally is given by a
trigonometric function, since we are interested in a spatial
hypersurface of constant time. For the case where the other hole is
spinning, Eq.~(\ref{approx-z}) will contain corrections proportional
to $\chi$, but these terms will not change the overall scale of the
tidal tensors.

Throughout the rest of the paper we will generate plots of physical
quantities, such as $\psi_0$, $\Psi$ and $h_{ab}$. For the purpose of
plotting, we will have to make two choices: one regarding the physical
scenario that produces the perturbation of the external universe; and
another regarding the parameters of the background black hole. As for
the physical scenario, we will choose the external universe to be
given by another orbiting black hole in a quasicircular orbit. This
choice allows us to represent the tidal fields with the scaling given
in Eq.~(\ref{approx-z}). This scaling is not the exact functional form
of the tidal fields and, therefore, the plots generated will not be
exact. However, this scaling will allow us to provide plots accurate
enough to study the general features of the global structure of the
quantities plotted, as well as some local features near the horizons.
Regarding the parameter choice, we will assume $m_1=m_2$, $a = 0.99
m_1$, and $b=10 m_1$, where $M = m_1 + m_2 = 1$ is the total mass.
These choices are made in accordance with the slow-motion
approximation, while making sure the system is not inside its ISCO.
However, the formulae presented in this paper should apply to other
choices of physical scenarios and background parameters as well, as
long as these do not conflict with the slow motion approximation.

Given the chosen physical scenario, the formulae in this paper should
apply to other mass ratios and separations, as long as the orbital
velocity does not become too large. For the background parameters
chosen, the orbital velocity is approximately $V \approx 0.3$, which
indicates that, for fixed masses, we cannot reduce the orbital
separation by much more without breaking the slow-motion
approximation. However, since we provide explicit analytic formulae
for all relevant physical quantities, we can estimate their error by
considering the uncontrolled remainders, {\textit{i.e.}}\, the
neglected terms in the approximation. In particular, in a later
section, we will see that the uncontrolled remainders are still much
smaller than the perturbation itself even at $b=10 M$ as long as we
restrict ourselves to field points sufficiently near the outer horizon
of the background hole.

\begin{figure}
\includegraphics[angle=0,scale=0.33]{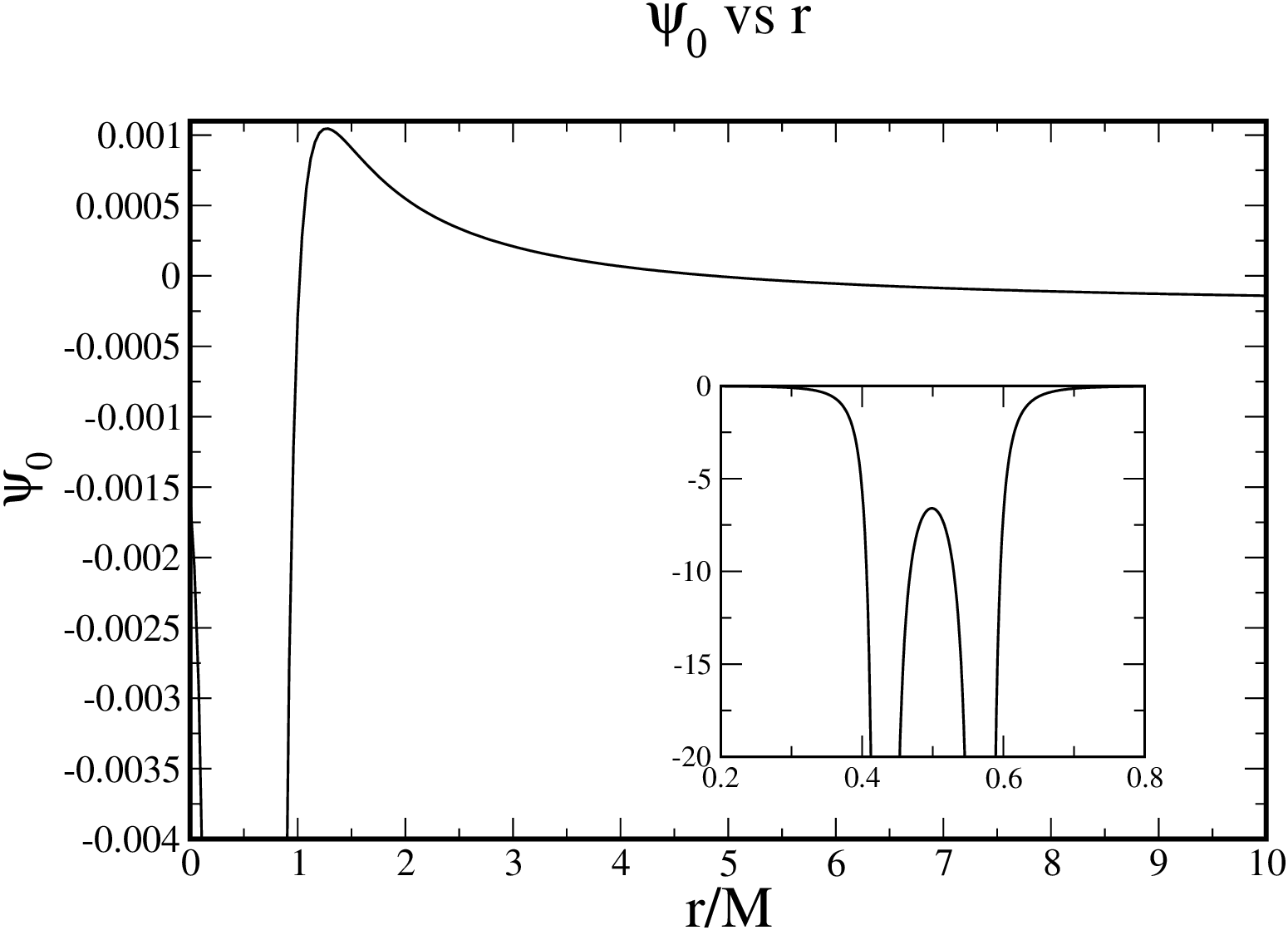}
\caption{\label{psi0-fig} Plot of the real part of the NP scalar $\psi_0$ along the
  $x$-axis with the plotting parameters described in Sec.~\ref{eric}.}
\end{figure}

In Fig.\ref{psi0-fig} we plot the real part of $\psi_0$ with the
plotting choices described earlier.
Observe that as $r$ becomes large the scalar asymptotes to a constant
given by ${\tilde{\psi}}_0$. Also observe that the functional behavior
of the scalar is drastically different as $r$ becomes small. In this
figure, as well as in future figures, we have chosen to include an
inset where we zoom to a region close to the horizons, so that we can
observe its local and global behavior. For the orbital parameter
chosen, the inner and outer horizons are given by $r_- \approx 0.43 M$
and $r_+ \approx 0.57 M$.  Observe from the inset in
Fig.~\ref{psi0-fig} that the scalar diverges at the horizons and as $r
\to 0$, which is due to the choice of tetrad. Finally, observe that,
except for where it diverges, the real part of $\psi_0$ is of $O(V^2)$
in the entire $3$-manifold.

\section{The potential}
\label{ori}
In this section we will use $\psi_0$ to construct the potential $\Psi$
by acting some differential operators on the NP scalar. This potential
must satisfy the vacuum Teukolsky equation for the $s=-2$ mode
together with the differential equation
\be
\label{def-potential}
\psi_0 = D^4[\bar{\Psi}],
\ee
where the overbar stands for complex conjugation and where the
differential operator is given by $D=l^a\partial_a$
\cite{Wald:1978vm,Lousto:2002em,Ori:2002uv}.

Ori \cite{Ori:2002uv} has shown that the above differential equations
can be inverted with use of the Teukolsky-Starobinsky relation to obtain
\be
\label{potential-def}
\Psi = \frac{1}{p} \Delta^2 (D^{\dag})^4[\Delta^2 \bar{\psi}_0],
\ee
where here $\Delta$ is given by
\be
\label{delta}
\Delta = r^2 - 2 m_1 r + a^2,
\ee
while $p$ is a constant that for the time-independent case reduces to
\be
p = \left[l (l + 1) - s^2 + |s|\right]^2 \left[ l (l + 1) - s^2
    + |s| + 2 \right]^2.
\ee
In our case, since $s$ and $l$ refer to $\psi_0$, $s= 2$ and $l=2$ so
that this constant becomes $p=576$.

The differential operator $D^{\dag}$ is given in spherical
Brill-Lindquist (BL) coordinates $(t_{BL},r,\theta,\phi_{BL})$ by
\be
\label{delta-def}
D^{\dag}_{BL} = \partial_r - \frac{a}{\Delta} \partial_{\phi_{BL}},
\ee
neglecting any time dependence, since time derivatives will only
contribute at a higher order. In order to compute $\Psi$ in Kerr
coordinates, we must transform the above differential operator. The
transformation between Kerr and BL coordinates is given by
\ba
dv &=& dt_{BL} + dr_{BL} \left( \frac{2 m_1 r}{\Delta} + 1\right),
\nonumber \\
dr &=& dr_{BL},
\nonumber \\
d\theta &=& d\theta_{BL},
\nonumber \\
d\phi &=& d\phi_{BL} + dr_{BL} \frac{a}{\Delta}. 
\ea
After transforming the differential operator $D^{\dag}_{BL}$ we obtain 
\be
D^{\dag} = \partial_r,
\ee
because the $dr_{BL}/d\phi$ term in the transformation cancels the
$\phi$ dependence. Note that $r$ and $\theta$ do not change in this
transformation and, thus, $\Delta$ remains unchanged. Also note that
$\Psi$ is a scalar constructed from differential operators on $\psi_0$
and, since the latter is a scalar, $\Psi$ will also be gauge
invariant.

Before plugging in Eq.~(\ref{psi0}) into Eq.~(\ref{potential-def}) to
compute $\Psi$, let us try to simplify these expressions. In the
previous section, we defined the inner and outer horizons $r_+$ and
$r_-$, as well as the new variable $x$. We can invert the definition
of $x$ so that it becomes a definition for $r$ as a function of $x$
and then insert this into Eq.~(\ref{delta}). We then obtain
\be
\Delta = 4 \eta x (1 + x),
\ee
where we have defined $\eta = m_1^2 - a^2$. It is clear now that when we
combine the square of this expression with Eq.~(\ref{psi0}) some
cancellations will occur that will simplify all future calculations.

We are now ready to compute $\Psi$, but first let us rewrite the
function we want to differentiate, namely
\be
\Delta^2 \bar{\psi}_0 = \sum_{m \neq 0} \bar{C}_m \bar{F}(x) \;
_2{\bar{Y}}_2^m(\theta,\phi_{BL}), 
\ee
where $F(x)$ is shorthand for the aforementioned hypergeometric function
and where $C_m$ is a new function of advanced time only given by 
\be
C_m = - 16 B_m \eta^2 = - 16 A_m \eta^2 z_m(v).
\ee
Note that the angular dependence occurs in the spherical harmonics,
while the only $x$ dependence is in the hypergeometric function. We
can transform the $D^{\dag}$ operator to $x$ space to obtain
\be
D^{\dag} = \partial_r  = \frac{1}{2 \eta^{1/2}} \partial_x.
\ee

Applying all these simplifications, $\Psi$ becomes
\be
\Psi = \frac{\Delta^2}{576} \sum_{m \neq 0} \bar{C}_m \;
_2{\bar{Y}}_2^m(\theta,\phi) \frac{1}{16 \eta^2} \partial_x^4[\bar{F}(x)].
\ee
We can now apply all derivatives to obtain 
\be
\Psi = \frac{\Delta^2}{576} \sum_{m \neq 0} \bar{C}_m \;
_2{\bar{Y}}_2^m(\theta,\phi) \frac{\bar{F}^{(4)}}{16 \eta^2}, 
\ee
where we have used the shorthand $\bar{F}^{(n)}$ which stands for the
$n$th derivative of the complex conjugate of the hypergeometric
function. This derivative is given by 
\be 
\label{decomp-F}
\bar{F}^{(4)} = \frac{288}{(2 i m  \gamma + 1)i m \gamma(-2 i m \gamma +
    1) (-2 i m \gamma + 2)},
\ee
Note that we can reexpress the constant $\gamma$ in terms of $\eta$ as
$\gamma = a/(2 \eta^{1/2})$. Interestingly this constant combines with
$\bar{C}_m$ to return an overall constant that is purely real so that
$\Psi$ is given by
\be
\label{potential}
\Psi = - \frac{1}{24} \Delta^2 \sum_{m \neq 0} Y^m(\theta) e^{-
  i m \phi} \bar{z}_m(v),
\ee
where here $Y^m$ stands for the $l=2$ spherical harmonics with zero
$\phi$ dependence. We have checked that the potential of
Eq.~(\ref{potential}) indeed satisfies the definition of
Eq.~(\ref{def-potential}) as well as the Teukolsky equation for the
$s=-2$ mode with angular eigenvalue $E=10$. Note that $\Psi$ has units
of mass squared because the electric and magnetic tidal fields scale
as the inverse of the mass squared.  Furthermore, note that eventhough
$\psi_0$ is singular at the horizon, $\Psi$ is finite and actually
vanishes there.

Next, we proceed to decompose $\Psi$ into real and imaginary parts.
The potential contains $2$ complex terms, namely the $\phi$ part of
the spherical harmonics and the electric and magnetic tidal tensors.
Decomposing $\Psi$ we obtain
\begin{align}
\label{final-potential}
\Psi_{R}  &= - \frac{\Delta^2}{24} \sum_{m \neq 0}
  \; Y^m \left( \Re(z_{m}) \cos{m \phi} -
  \Im(z_{m}) \sin{m \phi} \right), 
\nonumber \\
\Psi_{I}  &= +\frac{\Delta^2}{24} \sum_{m \neq 0}
  \; Y^m \left( \Re(z_{m}) \sin{m \phi} + \Im(z_{m}) \cos{m \phi}
  \right)\,,
\end{align}
where recall that $(\alpha_{m},\beta_{m})$ in $z_{m}$ are complex functions
of $v$. Note that the entire radial dependence is encoded in $\Delta^2$,
whereas the angular dependence is hidden in the spherical harmonics.
The time dependence occurs only in the the tidal fields that should be
determined via asymptotic matching, as mentioned previously. This is
the potential in Kerr coordinates associated with the $\psi_0$
calculated in the previous section in the slow-motion approximation.

\begin{figure}
\includegraphics[angle=0,scale=0.33]{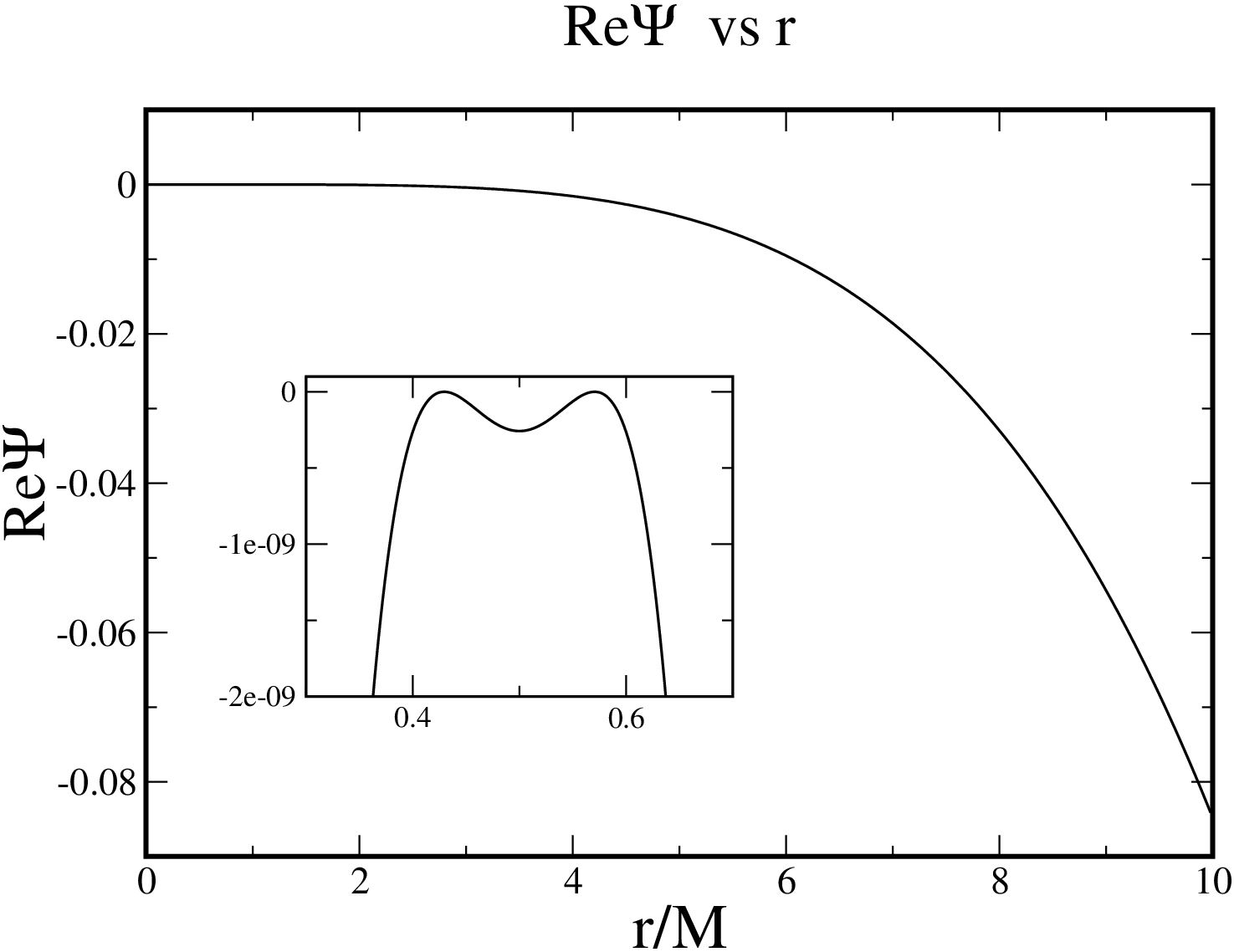}
\caption{\label{Psi} Plot of the potential $\Psi_R$ along the
$x$-axis with the plotting parameters described in Sec.~\ref{eric}.}
\end{figure}

In Fig.~\ref{Psi} we plot $\Psi_R$ with the plotting choices described
in Sec.~\ref{eric}.
Observe from the inset that the potential has nodes at both horizons.
Also observe that the potential does not asymptote to a constant, but
instead it grows quartically. This growth is due to the factor of
$\Delta^2$ that dominates at large radius. Finally, note that the
potential is still of $O(V^2)$ for radii sufficiently close to the
hole (roughly for $r < 8 m$).

\section{The metric perturbation}
\label{lousto}
In this section we compute the metric perturbation by applying
Chrzanowski's differential operator to the potential calculated in the
previous section. The full metric of the spacetime is given by 
\be
g_{a b} = g^B_{a b} + h_{a b},
\ee
where $g^B_{a b}$ is the background metric and $h_{a b}$ is the
perturbation. Since $\psi_0$ and $\Psi$ were computed in Kerr
coordinates, the background should also be in this coordinate system.
This background is given then by
\ba
g^B_{00} &=& -\left(1 - \frac{2 m_1 r}{\Sigma}\right),
\nonumber \\
g^B_{01} &=& 1,
\nonumber \\
g^B_{03} &=& -\frac{m_1 r}{\Sigma} (2 a \sin^2{\theta}),
\nonumber \\
g^B_{13} &=& -a \sin^2{\theta},
\nonumber \\
g^B_{22} &=& \Sigma,
\nonumber \\
g^B_{33} &=& (r^2 + a^2) \sin^2{\theta} + \frac{2 m_1 r}{\Sigma} (a^2 \sin^4{\theta}),
\ea
where $m_1$ is the mass of the background black hole, $a$ is its spin
parameter, related to the angular momentum vector by $\vec{S} = m
\vec{a}$, and where $\Sigma = r^2 + a^2 \cos^2{\theta}$.

Let us now construct the metric perturbation. We will work with the
form of the differential operator presented in
Ref.~\cite{Campanelli:1998jv}, namely
\ba
\label{MP}
h_{a b} &=& 2 \Re \left( \left\{ -l_{a} l_{b} ( \delta + \bar{\alpha} + 3
\beta - \tau)(\delta + 4 \beta + 3 \tau) 
\right. \right. 
\\ \nonumber
&& \left. \left. 
- m_{a} m_{b}(D - \rho)(D + 3 \rho) + l_{(a} m_{b)} \left[(D-2
  i \rho_I)
\right. \right. \right. 
\\ \nonumber 
&& \left. \left. \left. 
(\delta + 4 \beta + 3 \tau) + ( \delta + 3 \beta -
  \bar{\alpha} - \bar{\pi}  - \tau)
\right. \right. \right.
\\ \nonumber 
&& \left. \left. \left. 
(D + 3 \rho)\right] \right\} \Psi \right), 
\ea
where we have replaced $\bar{\rho} - \rho = -2 i \rho_I$. The metric
perturbation constructed in this fashion will be in the ingoing
radiation gauge (IRG), which is defined by the conditions
\be
\label{IRG}
h_{ll} = h_{ln} = h_{lm} = h_{l\bar{m}}=h_{m\bar{m}} = 0.
\ee

In Eq.~(\ref{MP}) there are terms that depend on the differential
operators $D = l^{a} \partial_{a}$ and $\delta = m^{a} \partial_{a}$,
which in turn depends on the tetrad. To be consistent, we will
continue to work with the Kinnersly tetrad in Kerr coordinates given
by
\ba
e_{(1)}^{a} &=& l^{a} = \left( 2 \frac{r^2 + a^2}{\Delta}, 1, 0 \frac{2
    a}{\Delta} \right),
\nonumber \\
e_{(2)}^{a} &=& n^{a} = \left( 0,-\frac{\Delta}{2 \Sigma}, 0, 0 \right). 
\nonumber \\
e_{(3)}^{a} &=& m^{a} = \frac{1}{\sqrt{2} (r + i a \cos{\theta})} \left[ i
  a \sin{\theta}, 0, 1,
\frac{i}{\sin{\theta}} \right],
\nonumber \\
e_{(4)}^{a} &=& \bar{m}^{a},
\ea
where the overbar stands for complex conjugation. Note that the
$m^{a}$ vector is the same as the one in BL coordinates, but $l^{a}$
and $n^{a}$ are different. The differential operators associated with
this tetrad in Kerr coordinates are
\ba
D &=& \partial_r + \frac{2a}{\Delta} \partial_{\phi}
\nonumber \\
\delta &=& \frac{1}{\sqrt{2}(r + i a \cos{\theta})}
\left(\partial_{\theta} + \frac{i}{\sin{\theta}}
\partial_{\phi} \right),
\ea
where once more we neglect the time derivatives by use of the slow
motion approximation. The covariant form of the tetrad in these
coordinates is given by
\ba
l_a &=& l_{a}^R = \left( 1 , -\frac{2 \Sigma}{\Delta}, 0, -a 
\sin^2{\theta}\right),
\\ \nonumber
n_a &=& n_{a}^R =  \frac{1}{2 \Sigma} \left(\Delta, 0, 0, -a
  \Delta \sin^2{\theta}\right),
\nonumber \\
m_{a}^R &=& \frac{a \sin{\theta}}{\sqrt{2} \Sigma}
\left[a \cos{\theta}, 0, -\frac{r \Sigma}{a \sin{\theta}}, - 
  \cos{\theta} (r^2+a^2)\right],
\nonumber \\ \nonumber
m_{a}^I &=& \frac{1}{\sqrt{2} \Sigma}
\left[a r \sin{\theta}, 0, a \Sigma \cos{\theta}, - r \sin{\theta} (r^2 + a^2) \right],
\ea
where the superscript $I$ and $R$ stand for the imaginary and real
parts respectively. One can show that if this tetrad is used we can
recover the background metric $g^B_{ab}$ with the formula $g^B_{a b} =
2 l_{(a} n_{b)} + 2 m_{(a} \bar{m}_{b)}$.

Eq.~(\ref{MP}) contains terms that depend on the spin coefficients of
the background \cite{Chandra}. These coefficients, also called Ricci
rotation coefficients for the case where the tetrad is non-null, are
simply contraction of the tetrad with its derivatives. In the tetrad
formalism, these quantities can also be related to the Riemann tensor.
Let us decompose the spin coefficients into real and imaginary parts,
{\textit{i.e.}}
\ba
\label{spin-coeff-decomp}
\rho_R &=& -\frac {r}{\Sigma},
\qquad 
\rho_I = -\frac {a}{\Sigma} \cos{\theta},
\nonumber \\ 
\beta_R &=& \frac{\sqrt{2}}{4} \frac{r}{\Sigma} \cot{\theta},
\qquad 
\beta_I = -\frac{\sqrt{2}}{4} \frac{a}{\Sigma} \cot{\theta}
\cos{\theta},
\nonumber \\
\pi_R &=& -\sqrt{2} \frac{a^2}{\Sigma^2} r \sin{\theta}\cos{\theta},
\nonumber \\
\pi_I &=& \frac{\sqrt{2}}{2} \frac{a}{\Sigma^2} \left(r^2 - a^2
  \cos^2{\theta}\right) \sin{\theta},
\quad
\tau_I = - \frac{\sqrt{2}}{2} \frac{a}{\Sigma} \sin{\theta},
\nonumber \\ 
\alpha_R &=& -\sqrt{2} \frac{a^2}{\Sigma^2} r \sin{\theta}\cos{\theta}
- \frac{\sqrt{2}}{4} \frac{r}{\Sigma} \cot{\theta},
\nonumber \\
\alpha_I &=& \frac{\sqrt{2}}{2} \frac{a}{\Sigma^2}
\sin{\theta}\left(r^2 - a^2 \cos^2{\theta}\right) - \frac{\sqrt{2}}{4}
\frac{a}{\Sigma} \cot{\theta} \cos{\theta}.
\nonumber \\
\ea
These spin coefficients are the same as those obtained with the
Kinnersly tetrad in BL coordinates. This invariance is due to the fact
that the spin coefficients are only tetrad dependent but still gauge
invariant. 

We will split Eq.~(\ref{MP}) into $4$ terms in order to make
calculations more tractable. The split is as follows: the term
proportional to $l_{a} l_{b}$ will be denoted term $A$; the term
proportional to $m_{a} m_{b}$ will be referred to as term $B$; the
first half of the term proportional to $l_{(a} m_{b)}$ will be
denoted term $C$; and the remaining of this term will be referred to
as term $D$. In this manner we have
\ba
h_{a b}^A &=& \left[ -l_{a} l_{b} ( \delta + \bar{\alpha} + 3
\beta - \tau)(\delta + 4 \beta + 3 \tau) \right] \Psi,
\nonumber \\
h_{a b}^B &=& \left[- m_{a} m_{b}(D - \rho)(D +
3 \rho) \right] \Psi,
\nonumber \\
h_{a b}^C &=& \left[l_{(a} m_{b)} (D-2 i \rho_I)(\delta + 4
    \beta + 3 \tau) \right] \Psi,
\nonumber \\
h_{a b}^D &=& \left[l_{(a} m_{b)}( \delta + 3 \beta - \bar{\alpha} - \bar{\pi}
    - \tau)(D + 3 \rho) \right] \Psi,
\ea
and the full metric perturbation is given by
\be
h_{a b} = 2 \Re ( h_{a b}^A + h_{a b}^B + h_{a b}^C +
h_{a b}^D ). 
\ee

With this split we can now proceed to simplify each expression more
easily. One such simplification is to operate with the differential
operators. Expanding all terms we obtain 
\begin{widetext}
\ba
\label{split-pi-op}
h_{a b}^A &=& - l_{a} l_{b} \left[ \delta^2 \Psi +
  \Psi \left( 4 \delta \beta + 3 \delta \tau \right) + \left( 7 \beta
   + 2 \tau  + \bar{\alpha} \right) \delta \Psi + \left( \bar{\alpha} +
     3 \beta - \tau \right) \left(4 \beta + 3 \tau \right) \Psi
 \right],
\nonumber \\
h_{a b}^B &=& - m_{a} m_{b} \left[ D^2 \Psi + 2
  \rho D \Psi + 3 \Psi D \rho - 3 \rho^2 \Psi \right],
\nonumber \\ 
h_{a b}^C &=& l_{(a} m_{b)} \left[D \delta\Psi + 4
  \Psi D\beta + 3 \Psi D \tau + \left( 4 \beta + 3 \tau \right) D\Psi
  -2 i \rho_I \left( 4 \beta + 3 \tau\right)
  \Psi - 2 i \rho_I \delta \Psi \right],
\nonumber \\ 
h_{a b}^D &=& l_{(a} m_{b)} \left[ \delta D \Psi +
  3 \Psi \delta \rho + 3 \rho \delta \Psi + \left( 3 \beta - \bar{\pi}
    - \tau - \bar{\alpha} \right) D \Psi + \left( 3 \beta - \bar{\pi}
    - \tau - \bar{\alpha} \right) 3 \rho \Psi\right],
\ea
\end{widetext}
These expressions give the metric perturbation in terms of the action
of the differential operators on the spin coefficients and the
potential.

\subsection{Action of the Differential Operators}
In order to provide explicit formulae for the metric perturbation we
must investigate how the differential operators act on the spin
coefficients and on the potential. Let us first concentrate on the
action of the differential operators on the spin coefficients. After
taking the necessary derivatives and decomposing the result into
imaginary and real parts, we obtain
\ba
\label{diff-op-on-spin-coeff}
D\rho &=& D_{\rho,R} + i D_{\rho,I} = 
\frac{{r}^{2} -{a}^{2} \cos^2{\theta}}{\Sigma^2}
+ i \left[ 2 \frac{r a}{\Sigma^2} \cos{\theta} \right],
\nonumber \\
D\tau &=& i D_{\tau,I} = i \sqrt{2} \frac{a r}{\Sigma^2} \sin{\theta},
\nonumber \\
D\beta &=& D_{\beta,R} + i D_{\beta,I} = -\frac{\sqrt{2}}{4} \frac{
  \cot{\theta}\left(r^2 - a^2 \cos^2{\theta}\right)}{\Sigma^2} 
\nonumber \\
&&  + i \left[ \frac{\sqrt{2}}{2} \frac{r a}{\Sigma^2} \cot{\theta}
   \cos{\theta}\right],
\nonumber \\
\delta \rho &=& \delta_{\rho,R} + i \delta_{\rho,I} =
-\frac{\sqrt{2}}{2} \frac{a^2}{\Sigma^2} \sin{\theta} \cos{\theta} 
\nonumber \\
&& + i \left[ \frac{\sqrt{2}}{2} \frac{a r}{\Sigma^2} \sin{\theta} \right],
\nonumber \\
\delta \beta &=& \delta_{\beta,R} + i \delta_{\beta,I} 
\nonumber \\
&=& \frac{a^4 \cos^6{\theta} - r^4 + 3 r^2 a^2 \cos^2{\theta} - 3
  \cos^4{\theta} r^2 a^2}{4 \Sigma^3 \sin^2{\theta}} 
\nonumber \\
&+& i \left[ r a \cos{\theta} \frac{3 a^2 \cos^4{\theta}
 - r^2 \cos^2{\theta} + 3 r^2 } {4 \sin^2{\theta}
 \Sigma^3}
\right.
\nonumber \\
&& \left. - r a \cos{\theta} \frac{a^2 \cos^2{\theta}}{4 \sin^2{\theta}
  \Sigma^3} \right],
\nonumber  \\
\delta \tau &=& \delta_{\tau,R} + i \delta_{\tau,I} = 
\frac{1}{2} a^2 \cos^2{\theta} \frac{-r^2 + a^2 \cos^2{\theta} 
 - 2 a^2} {\Sigma^3} 
\nonumber \\
&& + i \left[ \frac{1}{2} r a \cos{\theta} \frac{ -r^2 + a^2 \cos^2{\theta} 
 - 2 a^2} {\Sigma^3} \right].
\ea

We now need to act with the differential operators on the potential
itself. We can separate the $\phi$ dependence from these operators to
obtain
\ba
D \Psi &=& D_m \Psi,
\nonumber \\
\delta \Psi &=& \delta_m \Psi,
\nonumber \\
(D \delta) \Psi &=& (D\delta)_m \Psi,
\nonumber \\
(\delta D) \Psi &=& (\delta D)_m \Psi
\ea
where 
\ba
D_m &=& \partial_r - 2 i \frac{ma}{\Delta} =
\frac{1}{2 \eta^{1/2}} \partial_x - \frac{ i m a}{2 \eta} \frac{1}{x
  (1 + x)},
\nonumber \\
\delta_m &=& \delta_0 \left(\partial_{\theta} + \frac{m}{\sin{\theta}}
  \right),
\nonumber \\
\delta_0 &=& \frac{-\bar{\rho}}{\sqrt{2}}, \qquad \delta_{0,r} =
-\sqrt{2} \delta_0^2.
\ea
We can also compute the square of these operators acting on $\Delta^2
Y^m$, where $Y^m$ stands for the spherical harmonics with no $\phi$
dependence. Doing so we obtain
\ba
&& D_m^2[\Delta^2] = 8 (r-m_1)^2 + 4 \Delta - 4 m^2 a^2 
\nonumber \\
&& \qquad \qquad - 12 i m a(r-m_1),
\nonumber \\
&& \delta_m^2[Y^m] = \delta_0^2 Y^m_{,\theta\theta} 
\nonumber \\
&& + \left[ \delta_0 \left(\frac{i a \sin{\theta}}{\sqrt{2} \Sigma} +
    \frac{\delta_0}{\Sigma} a^2 \sin{2 \theta} \right) 
+ 2 \delta_0^2 \frac{m}{\sin{\theta}} \right] Y^m_{,\theta} 
\nonumber \\
&& + \left[\frac{\delta_0 m}{\sin{\theta}} \left(  \frac{i a
      \sin{\theta}}{\sqrt{2} \Sigma} + \frac{\delta_0}{\Sigma} a^2
    \sin{2 \theta} \right) 
\right. 
\nonumber \\
&& \left.  + \frac{\delta_0^2 m}{\sin^2{\theta}} \left( m -
    \cos{\theta}\right) \right] Y^m, 
\nonumber \\
&& (\delta D)_m[\Delta^2 Y^m] = \delta_0 \left( Y^m_{,\theta} +
  \frac{m}{\sin{\theta}} Y^m\right) \left[ 4 \Delta (r-m_1) 
\right.
\nonumber \\
&& \left. \qquad \qquad \qquad - 2 i m a \Delta\right],
\nonumber \\
&& (D \delta)_m[\Delta^2 Y^m] = -\sqrt{2} \delta_0^2 \Delta^2 \left( Y^m_{,\theta} +
  \frac{m}{\sin{\theta}} Y^m \right) 
\nonumber \\
&& \qquad \qquad \qquad + (\delta D)_m[\Delta^2 Y^m],
\ea
where the commas stand for partial differentiation. 

We now have all the ingredients to compute the action of the
differential operators on the potential. Doing so we obtain 
\ba
\label{diff-op-on-psi}
D_m \Psi &=& - \frac{\Delta}{12} \sum_{m \neq 0} \; Y^m \bar{z}_m
e^{-im \phi} \left[2 (r-m_1) - i m a \right], 
\nonumber \\
\delta_m \Psi &=& - \frac{\Delta^2}{24} \sum_{m \neq 0} \left(
  Y^m_{,\theta} + \frac{m}{\sin{\theta}} Y^m \right) \; e^{-im \phi} 
\bar{z}_m  \delta_0, 
\nonumber \\
(D_m)^2 \Psi &=& - \frac{1}{6} \sum_{m \neq 0} \; Y^m 
e^{-im \phi} \bar{z}_m \left[  2 (r-m_1)^2 + \Delta 
\right. 
\nonumber \\
&& \left. -  m^2  a^2 - 3 i m a(r-m_1) \right],
\nonumber \\
(\delta_m)^2 \Psi &=& - \frac{\Delta^2}{24} \sum_{m \neq 0} \; 
e^{-im \phi} \bar{z}_m \left\{ \delta_0^2 Y^m_{,\theta \theta} +
  \left[ \delta_0 \frac{i a \sin{\theta}}{\sqrt{2} \Sigma} 
\right. \right.
\nonumber \\
&& \left. \left. 
+ \delta_0^2 \left( \frac{a^2 \sin{2 \theta}}{\Sigma} + 2
  \frac{m}{\sin{\theta}} \right) \right] Y^m_{,\theta} +
\left[\frac{\delta_0 m}{\sin{\theta}} 
\right. \right. 
\nonumber \\
&& \left. \left. 
\left( \frac{i a \sin{\theta}}{\sqrt{2} \Sigma} 
+ \frac{\delta_0}{\Sigma} a^2 \sin{2 \theta} \right) +
\frac{\delta_0^2 m}{\sin^2{\theta}} 
\left( m 
\right. \right. \right. 
\nonumber \\
&& \left. \left. \left. 
- \cos{\theta}\right) \right] Y^m \right\},
\nonumber \\
(\delta D)_m \Psi &=& -\frac{\Delta}{12} \sum_{m \neq 0} \; 
\left( Y^m_{,\theta} + \frac{m}{\sin{\theta}} Y^m\right) 
e^{-im \phi} \bar{z}_m 
\nonumber \\
&& \left[ 2 \delta_0 (r-m_1) -  \delta_0 i m a \right],
\nonumber \\
(D \delta)_m \Psi &=&  (\delta D)_m \Psi + \frac{\sqrt{2}}{24}
\Delta^2 \sum_{m \neq 0} \left( Y^m_{,\theta} +
  \frac{m}{\sin{\theta}} Y^m \right) 
\nonumber \\
&& e^{-im \phi} \bar{z}_m \delta_0^2,
\ea

In order to complete the calculation, we need to provide explicit
formulae for the first and second derivatives of the spherical
harmonics. These derivatives are given by
\ba
Y^{\pm 1}_{,\theta} &=& -2 \cos^2{\theta} + 1 \pm \cos{\theta},
\nonumber \\
Y^{\pm 1}_{,\theta\theta} &=& \sin{\theta} \left(4 \cos{\theta} \mp 1\right),
\nonumber \\
Y^{\pm 2}_{,\theta} &=& -\frac{1}{2} \sin{\theta} \left( \cos{\theta}
  \mp 1\right),
\nonumber \\
Y^{\pm 2}_{,\theta\theta} &=& - \cos^2{\theta} \pm \frac{1}{2}
\cos{\theta} + \frac{1}{2}.
\ea
Note that the spherical harmonics and all of its derivatives are
purely real. 

\subsection{Decomposition into real and imaginary parts}
We will conclude this section by explicitly taking the real part of
the metric perturbation, so as to have explicit formulae for the
metric in terms of only the real and imaginary parts of the spin
coefficients, the potential and the action of the differential
operators on these quantities.

Before decomposing Eq.~(\ref{split-pi-op}), however, we must decompose
the action of the differential operators on the potential,
{\textit{i.e.}}\ Eq.~(\ref{diff-op-on-psi}). Let us first note that
the action of any differential operator on the potential always
contains the product of $3$ complex terms, the first $2$ of which are
always $e^{-im\phi}$ and $\bar{z}_m$. The third term varies depending
on the differential operator. Let us define the third term with
superscripts as
\ba
c^{(D)} &=& - \frac{\Delta}{12} \sum_{m \neq 0} \; Y^m 
\left[2 (r-m_1) - i m a \right], 
\nonumber \\
c^{(\delta)} &=& - \frac{\Delta^2}{24} \sum_{m \neq 0} \left(
  Y^m_{,\theta} + \frac{m}{\sin{\theta}} Y^m \right) \; \delta_0, 
\nonumber \\
c^{(D^2)} &=& - \frac{1}{6} \sum_{m \neq 0} \; Y^m 
\left[  2 (r-m_1)^2 + \Delta 
\right. 
\nonumber \\
&& \left. -  m^2  a^2 - 3 i m a(r-m_1) \right],
\nonumber \\
c^{(\delta)^2} &=& - \frac{\Delta^2}{24} \sum_{m \neq 0} \; 
\left\{ \delta_0^2 Y^m_{,\theta \theta} +
\left[ \delta_0 \frac{i a \sin{\theta}}{\sqrt{2} \Sigma} 
\right. \right.
\nonumber \\
&& \left. \left. 
+ \delta_0^2 \left( \frac{a^2 \sin{2 \theta}}{\Sigma} + 2
  \frac{m}{\sin{\theta}} \right) \right] Y^m_{,\theta} +
\left[\frac{\delta_0 m}{\sin{\theta}} 
\right. \right. 
\nonumber \\
&& \left. \left. 
\left( \frac{i a \sin{\theta}}{\sqrt{2} \Sigma} 
+ \frac{\delta_0}{\Sigma} a^2 \sin{2 \theta} \right) +
\frac{\delta_0^2 m}{\sin^2{\theta}} 
\left( m 
\right. \right. \right. 
\nonumber \\
&& \left. \left. \left. 
- \cos{\theta}\right) \right] Y^m \right\},
\nonumber \\
c^{(\delta D)}  &=& -\frac{\Delta}{12} \sum_{m \neq 0} \; 
\left( Y^m_{,\theta} + \frac{m}{\sin{\theta}} Y^m\right) 
\nonumber \\
&& \left[ 2 \delta_0 (r-m_1) -  \delta_0 i m a \right],
\nonumber \\
c^{(D \delta)} &=&  c^{(\delta D)} + \frac{\sqrt{2}}{24}
\Delta^2 \sum_{m \neq 0} \left( Y^m_{,\theta} +
  \frac{m}{\sin{\theta}} Y^m \right) 
\nonumber \\
&& \delta_0^2,
\ea

In general, if we want to decompose the product of $3$ complex
quantities $a$, $b$ and $c$ we will obtain
\ba
\label{decomp-abc}
(abc)_R &=& c_R \left(a_R b_R - a_I b_I\right) - c_I \left( a_R b_I +
  a_I b_R \right), 
\nonumber \\
(abc)_I &=& c_R \left(a_R b_I + a_I b_R\right) + c_I \left( a_R b_R - a_I
  b_I\right).
\\ \nonumber
\ea
Since we have identified $a=e^{-im\phi}$ and $b=\bar{z}_m$, their real
and imaginary parts are $a_R = \cos{m \phi}$, $a_I = - \sin{m \phi}$,
$b_R = \alpha_m$ and $b_I = - \beta_m$. Finally, if we further
decompose $c$ we obtain
\ba
c^{(D)}_R &=& -\frac{\Delta}{6} \sum_{m \neq 0} Y^m (r - m_1),
\nonumber \\
c^{(D)}_I &=& \frac{\Delta}{12} \sum_{m \neq 0} Y^m m a,
\nonumber \\
c^{(\delta)}_R &=& -\frac{\Delta^2}{24} \sum_{m \neq 0} \left(
  Y^m_{,\theta} + \frac{m}{\sin{\theta}} Y^m \right) \delta_{0,R},
\nonumber \\
c^{(\delta)}_I &=& -\frac{\Delta^2}{24} \sum_{m \neq 0} \left(
  Y^m_{,\theta} + \frac{m}{\sin{\theta}} Y^m \right) \delta_{0,I},
\nonumber \\
c^{(D^2)}_R &=& -\frac{1}{6} \sum_{m \neq 0} Y^m \left[2\left(r -
    m_1\right) + \Delta - m^2 a^2 \right],
\nonumber \\
c^{(D^2)}_I &=& \frac{1}{2} \sum_{m \neq 0} Y^m ma \left(r -
    m_1\right), 
\nonumber \\
c^{(\delta D)}_R &=& -\frac{\Delta}{12} \sum_{m \neq 0} \left(
  Y^m_{,\theta} + \frac{m}{\sin{\theta}} Y^m \right) \left[ 2
  \delta_{0,R} (r - m_1) \right.
\nonumber \\
&& \left. + \delta_{0,I} m a \right]
\nonumber \\
c^{(\delta D)}_I &=& -\frac{\Delta}{12} \sum_{m \neq 0} \left(
  Y^m_{,\theta} + \frac{m}{\sin{\theta}} Y^m \right) \left[ 2
  \delta_{0,I} (r - m_1) \right. 
\nonumber \\
&& \left. - \delta_{0,R} m a \right]
\nonumber \\
c^{(D \delta)}_R &=& c^{(\delta D)}_R + \frac{\sqrt{2}}{24} \Delta^2
\sum_{m \neq 0} \left( Y^m_{,\theta} + \frac{m}{\sin{\theta}} Y^m
\right) \left(\delta_{0,R}^2 
\right.
\nonumber \\
&& \left. - \delta_{0,I}^2 \right) 
\nonumber \\ 
c^{(D \delta)}_I &=&  c^{(\delta D)}_I + \frac{\sqrt{2}}{12} \Delta^2
\sum_{m \neq 0} \left( Y^m_{,\theta} + \frac{m}{\sin{\theta}} Y^m
\right) \delta_{0,R} \delta_{0,I} \nonumber
\ea

\ba
\label{diff-ops-on-psi-decomped2}
c^{(\delta^2)}_R &=& -\frac{\Delta^2}{24} \sum_{m \neq 0} \left\{
  Y^m_{,\theta\theta} \left(\delta_{0,R}^2 - \delta_{0,I}^2 \right) +
  \left[ - \delta_{0,I} \frac{a \sin{\theta}}{\sqrt{2} \Sigma} 
\right. \right.
\nonumber \\
&& \left. \left. 
+ \left( \delta_{0,R}^2 -   \delta_{0,I}^2 \right) \left( \frac{a^2
  \sin{2 \theta}}{\Sigma} + \frac{2 m}{\sin{\theta}} \right) \right]
  Y^m_{,\theta} 
\right.  
\nonumber \\
&& \left.  
+ \left[ \left( \delta_{0,R}^2 - \delta_{0,I}^2 \right)  \left(\frac{m
  a^2 \sin{2\theta}}{\sin{\theta} \Sigma} + \frac{m}{\sin{\theta}^2}
  (m 
\right.\right.\right. 
\nonumber \\
&& \left. \Bigl. \left. 
- \cos{\theta}) \Bigr) - \delta_{0,I} \frac{m a}{\sqrt{2} \Sigma} 
  \right] Y^m \right\} 
\nonumber \\
c^{(\delta^2)}_I &=&  -\frac{\Delta^2}{24} \sum_{m \neq 0} \left\{
  Y^m_{,\theta\theta} 2 \delta_{0,R} \delta_{0,I} +
  \left[ \delta_{0,R} \frac{a \sin{\theta}}{\sqrt{2} \Sigma} 
\right. \right.
\nonumber \\
&& \left. \left. 
+ 2 \delta_{0,R} \delta_{0,I} \left( \frac{a^2
  \sin{2 \theta}}{\Sigma} + \frac{2 m}{\sin{\theta}} \right) \right]
  Y^m_{,\theta} 
\right.  
\nonumber \\
&& \left.  
+ \left[ 2 \delta_{0,R} \delta_{0,I}  \left(\frac{m
  a^2 \sin{2\theta}}{\sin{\theta} \Sigma} + \frac{m}{\sin{\theta}^2}
  (m 
\right.\right.\right.
\nonumber \\
&& \left. \Bigl. \left. 
- \cos{\theta}) \Bigr) + \delta_{0,R} \frac{m a}{\sqrt{2} \Sigma} 
  \right] Y^m \right\} 
\ea
and where
\be
\delta_{0,R} = \frac{r}{\sqrt{2} \Sigma} 
\qquad 
\delta_{0,I} = \frac{-a}{\sqrt{2} \Sigma} \cos{\theta}.
\ee
From these equations it is simple to reconstruct the real and
imaginary parts of the action of the differential operators on the
potential by combining
Eq.~(\ref{diff-ops-on-psi-decomped2})
and (\ref{decomp-abc}). For example, the real part of $D_m \Psi$ is
then given by
\ba
(D_m \Psi)_R &=& c_R^{(D)} \left(a_R b_R - a_I b_I\right) - c_I^{(D)}
\left( a_R b_I + a_I b_R \right) 
\nonumber \\
&=& \frac{\Delta}{12} \sum_{m \neq 0} Y^m \left[ -2 (r - m_1)
  \left(\Re(z_{m})(v) \cos{m \phi}   
\right. \right.
\nonumber \\
&-& \left.  \left. 
\Im(z_{m})(v) \sin{m \phi} \right) + ma \left (\Re(z_{m})(v) \sin{m \phi}
\right. \right. 
\nonumber \\
&+& \left. \left. 
 \Im(z_{m})(v) \cos{m \phi}  \right) \right]\,.
\ea
%


We are now finally ready to get a final expression for the metric
perturbation by taking the real part of Eq.~(\ref{split-pi-op}). Doing
so we obtain
\begin{widetext}
\ba
\label{MP1decomp}
\Re[h_{a b}^{(A)}] &=& - l_{a} l_{b} \left[ \Psi_R
  \left( 4 \delta_{\beta,R} + 4 \alpha_R \beta_R + 3 \alpha_I \tau_I +
    3 \tau_I^2 + 12 \beta_R^2 - 5 \beta_I \tau_I + 3 \delta_{\tau,R} +
    4 \alpha_I \beta_I - 12 \beta_I^2 \right) + \Psi_I \left( -24
    \beta_R \beta_I
\right. \right. 
\nonumber \\
&& \left. \left. 
 - 4 \delta_{\beta,I}  - 3 \delta_{\tau,I} - 5
    \tau_I \beta_R + 4 \alpha_I \beta_R - 3 \alpha_R \tau_I - 4
    \alpha_R \beta_I \right) + (\delta_m \Psi)_R \left( 7 \beta_R 
    + \alpha_R \right) + (\delta_m \Psi)_I \left( -7 \beta_I - 2
    \tau_I + \alpha_I \right) 
\right.
\nonumber \\
&& \left. + [(\delta_m)^2 \Psi]_R \right],
\nonumber \\ \nonumber \\
\Re[h_{a b}^{(B)}] &=& 2 [(D_m)^2 \Psi]_I m^R_{a} m^I_{b} +
[(D_m)^2 \Psi]_R (m^I_{a} m^I_{b} - m^R_{a} m^R_{b}) + (D_m \Psi)_R
\left[ 2 (m^I_{a} m^I_{b} -  m^R_{a} m^R_{b}) \rho_R + 4 m^R_{a}
m^I_{b} \rho_I \right]
\nonumber \\
&& 
 + (D_m \Psi)_I \left[4 m^R_{a} m^I_{b} \rho_R + 2 (  m^R_{a}
 m^R_{b} - m^I_{a} m^I_{b}) \rho_I \right]
+ 3 \Psi_R \left[ (m^I_{a} m^I_{b} - m^R_{a} m^R_{b}) (\rho_I^2 -
\rho_R^2) - 4 m^R_{a} m^I_{b} \rho_R \rho_I 
\right.
\nonumber \\
&& \left.
+ D_{\rho,R} (m^I_{a} m^I_{b} - m^R_{a} m^R_{b}) + 2 m^R_{a}
m^I_{b} D_{\rho,I}\right] + 6 \Psi_I \Bigl[ (m^I_{a} m^I_{b} - m^R_{a}
m^R_{b} ) \rho_R  \rho_I  + m^R_{a} m^I_{b} (\rho_I^2 -
\rho_R^2) 
\Bigr.
\nonumber \\
&& \left. + \frac{1}{2} (m^R_{a} m^R_{b} - m^I_{a} m^I_{b})
D_{\rho,I} + m^R_{a} m^I_{b} D_{\rho,R}\right],  
\nonumber \\ \nonumber \\
\Re[h_{a b}^{(C)}] &=& l_{(a} \left( m^R_{b)} \left\{ \left[(D
  \delta)_m \Psi\right]_R + 2 (\delta_m \Psi)_I \rho_I \right\} -
  m^I_{b)} \left\{ \left[(D \delta)_m \Psi\right]_I - 2 (\delta_m
  \Psi)_R \rho_I \right\} + (D_m \Psi)_R  \left[4 \beta_R m^R_{b)} -
  (4 \beta_I + 3 \tau_I) 
\right. \right.
\nonumber \\
&& \left. \left. 
 m^I_{b)} \right] - (D_m \Psi)_I \left[4 \beta_R m^I_{b)} + (4 \beta_I
  + 3 \tau_I) m^R_{b)} \right] + \Psi_R \left\{ \left[3 D_{\tau,R} +
  (8 \beta_I + 6 \tau_I) \rho_I + 4 D_{\beta,R}\right] m^R_{b)} + (-3
  D_{\tau,I} 
\right. \right.
\nonumber \\
&& \left. \left.
 - 4 D_{\beta,I} +  8 \rho_I \beta_R ) m^I_{b)} \right\} + \Psi_I
  \left[(-3  D_{\tau,I} - 4 D_{\beta,I} + 8 \rho_I \beta_R ) m^R_{b)}
  - [ 3  D_{\tau,R} + (8 \beta_I + 6 \tau_I) \rho_I + 4 D_{\beta,R}] 
  m^I_{b)}\right] \right)  
\nonumber \\ \nonumber \\
\Re[h_{a b}^{(D}] &=& l_{(a} \left\{ 3 (\delta_m \Psi)_R ( -
  m^I_{b)} \rho_I + m^R_{b)} \rho_R) - 3 (\delta_m \Psi)_I
  (m^I_{b)} \rho_R + m^R_{b)} \rho_I) + (D_m \Psi)_R \left[
  m^I_{b)} (-\alpha_I - \pi_I + \tau_I- 3 \beta_I) 
\right. \right.
\nonumber \\
&& \left. \left. 
+ m^R_{b)} (-\alpha_R + 3 \beta_R - \pi_R) \right] + (D_m \Psi)_I
  \left[ m^I_{b)} (-3 \beta_R + \pi_R + \alpha_R) + m^R_{b)} (-3
  \beta_I - \alpha_I + \tau_I - \pi_I) \right] + 3 \Psi_R 
\right. 
\nonumber \\
&&  \left. 
\left[ m^R_{b)} (-\pi_R  \rho_R + \delta_{\rho,R} - \alpha_R \rho_R -
  \alpha_I \rho_I + 3 \rho_R \beta_R -  \pi_I \rho_I +  \rho_I \tau_I
  - 3 \rho_I \beta_I ) +  m^I_{b)} (\alpha_R \rho_I -  \delta_{\rho,I}
  - 3 \beta_I \rho_R 
\right.\right.
\nonumber \\
&& \left. \left.
+ \pi_R \rho_I  -  \pi_I \rho_R - 3 \rho_I \beta_R +  \tau_I \rho_R -
  \alpha_I \rho_R ) \right] + 3 \Psi_I \left[m^R_{b)} (- \alpha_I
  \rho_R +  \pi_R \rho_I -  \delta_{\rho,I} -  \pi_I \rho_R +  \tau_I
  \rho_R -  3 \beta_I \rho_R  
\right. \right.
\nonumber \\
&& \left. \left. 
+  \alpha_R \rho_I  - 3 \rho_I \beta_R ) + m^I_{b)} (-\delta_{\rho,R}
  +  \alpha_I \rho_I +  \alpha_R \rho_R +  \pi_R \rho_R -  3\beta_R
  \rho_R + 3 \rho_I \beta_I -  \rho_I \tau_I +  \pi_I \rho_I) \right]
\right.
\nonumber \\
&& \left. 
 - m^I_{b)} \left[(\delta D)_m \Psi\right]_I  + m^R_{b)} \left[(\delta
   D)_m \Psi\right]_R \right\}. 
\ea
\end{widetext}

The full metric perturbation is then given by 
\be
\label{MP1}
h_{a b} = 2 \left[ \Re[h_{a b}^{(A)}] + \Re[h_{a b}^{(B)}] +
  \Re[h_{a b}^{(C)}] + \Re[h_{a b}^{(D)}] \right]. 
\ee
This is the metric of a tidally perturbed Kerr black hole in Kerr
coordinates. We can transform this metric to Kerr-Schild coordinates,
but this is left to the Appendix. We have checked that this metric
indeed satisfies the Einstein equations by linearizing the Ricci
tensor and verifying that all components vanish to first order.  We
have further checked that the metric perturbation is transverse and
traceless ($h^a_a =0$ and $h_{ab} l^a =0$) in the tetrad frame making
it suitable to study gravitational perturbations near the horizon.
Furthermore, we have checked that the conditions that define the IRG
[Eq.~(\ref{IRG})] are also satisfied. Another feature of this metric
is that its determinant is zero, which renders it non-invertible.
However, the full metric $g_{ab} = g_{ab}^B + h_{ab}$ is invertible
and, thus, the calculation of the Einstein tensor is straightforward.

The metric perturbation has now been expressed entirely in terms of
quantities explicitly defined in this paper. These quantities are the
real and imaginary parts of the spin coefficients, the potential and
the action of the differential operators on the spin coefficients and
the potential. The spin coefficients where decomposed in
Eq.~(\ref{spin-coeff-decomp}); the potential was decomposed in
Eq.~(\ref{final-potential}); the action of the differential operators
on the spin coefficients is given in
Eq.~(\ref{diff-op-on-spin-coeff}); and the action of these operators
on the potential is decomposed in
Eq.~(\ref{decomp-abc}) and (\ref{diff-ops-on-psi-decomped2}).

The metric perturbation possesses the general global features that it
diverges as $r \to \infty$ and it either diverges or converges to a
finite value as $r \to 0$. The behavior as $r \to \infty$ is to be
expected because the Chrzanowski procedure seizes to be valid far from
the hole. On the other hand, the behavior as $r \to 0$ is a bit more
surprising. In this region there are two different types of behavior:
either the perturbation remains finite or it diverges. These different
types of behavior depend on the component and axis we are
investigating. On the one hand, there are some components that either
are finite and of $O(V^2)$ or vanish as $r \to 0$ for all angles, such
as $h_{01}$, $h_{11}$, $h_{12}$, $h_{13}$ and $h_{22}$. On the other
hand, there are other components that diverge along certain axis as $r
\to 0$. For example, $h_{00}$, $h_{03}$ and $h_{33}$ diverge along the
$x$-axis, the $y$-axis and the $x$-$y$ diagonal, while $h_{02}$ and
$h_{23}$ diverge along the $y$-axis and $x$-$y$ diagonal. This
divergence is due to the choice of tetrad, since the fourth Kinnersly
tetrad vector clearly diverges as $r \to 0$. Note, however, that since
the divergences occur well inside the inner horizon they will be
causally disconnected with all physical processes ocurring outside the
outer horizon and, thus, these divergences are irrelevant to most
physical applications. This divergent behavior could nonetheless be
avoided if a different tetrad, such as the Hawking-Hartle one, is used
to compute the perturbation, but this will not be discussed here
further.

\begin{figure}
\includegraphics[angle=0,scale=0.33]{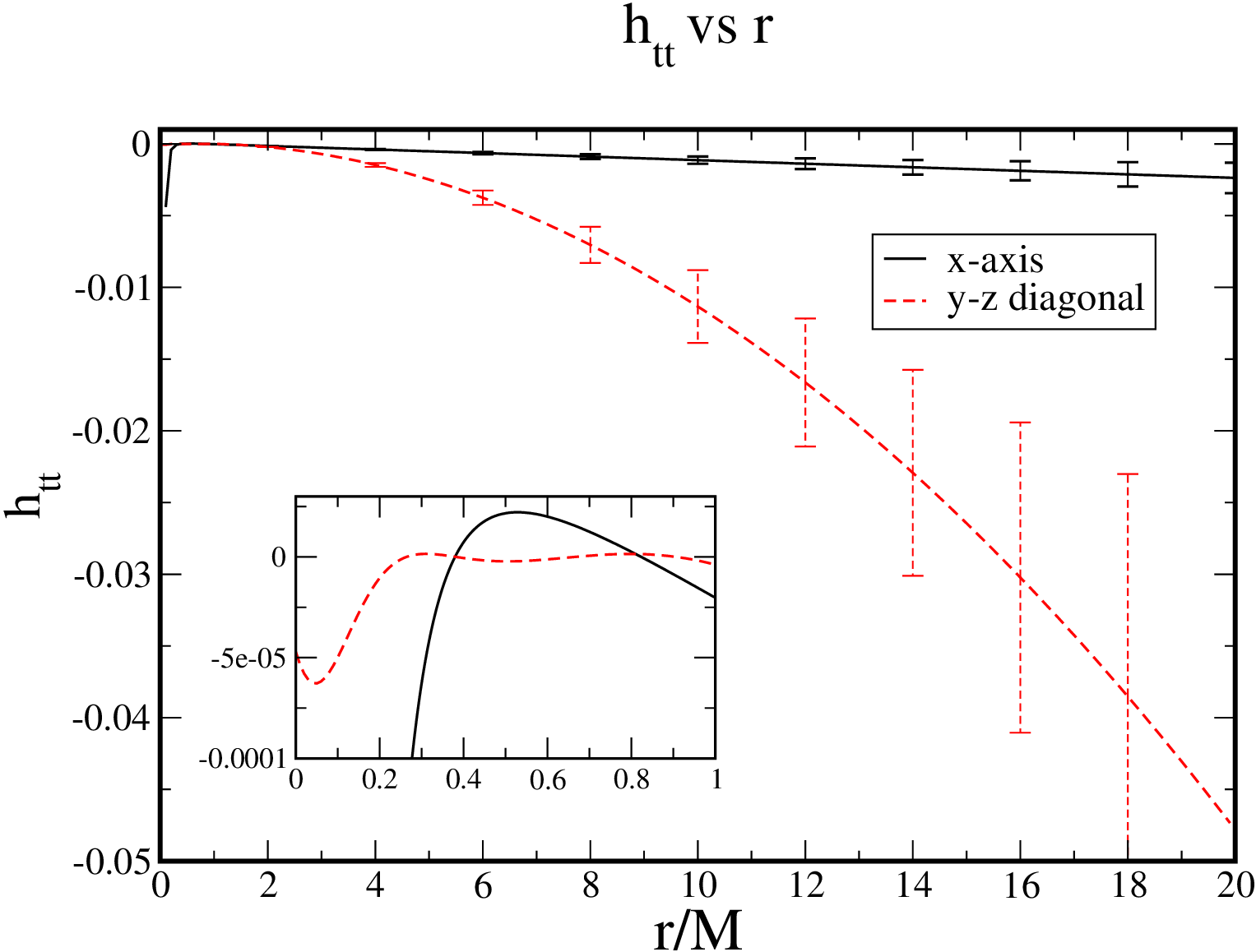}
\caption{\label{metric} Plot of the $00$ component of the metric
  perturbation along the $x$-axis (solid line) and along the $y$-$z$
  diagonal (dashed line) with the plotting parameters described in
  Sec.~\ref{eric}.}
\end{figure}

In order to illustrate this global behavior, we have plotted $h_{00}$
in Fig.~\ref{metric} with the plotting choices described in
Sec.~\ref{eric} along the $x$-axis and the $y$-$z$ diagonal ($\theta =
\pi/4$ and $\phi=\pi/2$).
Observe that the perturbation diverges as $r \to \infty$ and as $r \to
0$ along the $x$-axis, but it remains finite as $r \to 0$ along the
$y$-$z$ diagonal. Everywhere else, and in particular near the outer
horizon, the perturbation is of $O(V^2)$, where $V = (M/b)^{1/2}$ is
the orbital velocity. Finally, observe that the perturbation vanishes
close but not really at either horizon, remaining finite through them.

The divergence of the perturbation can be used to approximately
determine the region of validity of the approximation. As we have
discussed previously and seen in Fig.~\ref{metric}, the perturbation
will be valid inside a shell centered at the background hole. The
inner radius of this shell can be approximately determined by
graphically studying the radius where $h_{ab} > O(V^2)$, approximately
given by $r_{inner} \approx 0.01 M$. This value for the inner radius
of the shell is only an order of magnitude estimate, but is suffices
to see that $r_{inner} \ll r_-$. The causal structure of the region $r
< r_-$ is extremely complex and many of its features are known to be
unstable under small perturbations that destroy the symmetries of the
spacetime~\cite{Poisson}. However, note that the perturbation is well
behaved for $r-r_- \ll M$ and in particular it is of the order
predicted by the approximation [$O(V^2)$]. In any case, the region $r
< r_-$ is hidden inside the event horizon and most astrophysical
applications will be concerned with regions of small spacelike
separations from the outer horizon, and not the inner one.

The outer radius of the shell can be estimated by studying the
fractional error in the perturbation, which is determined by comparing
the perturbation to the uncontrolled remainders in the approximation.
The approximate error bars in Fig.~\ref{metric} are given by an
estimate of these uncontrolled remainders, $\delta h_{ab}$, which are
due to truncating the formal series solution at a finite order. In the
present case, this truncation is done at $O(V^2)$, where the tidal
fields provide this scaling. The uncontrolled remainders then will be
of $O(V^3)$ and should come from time derivatives of the tidal
tensors, {\textit{i.e.}}\ $\delta h_{ab} \propto \dot{h}_{ab}$. The
argument of the tidal tensors is $\omega v$, where $\omega=V/b$ is the
angular velocity and where $v$ is the advanced time coordinate. Any
time derivative will pull out a factor of $\omega$, which will in turn
increase the order of that term. However, in order for the
uncontrolled remainders to be dimensionally consistent ($\delta
h_{ab}$ must have the same dimensions as $h_{ab}$), we need to
multiply the time derivatives of the tidal tensors by $r$, so that
\be
\label{error}
\delta h_{ab} \approx r \dot{h}_{ab} \approx r (V/b) h_{ab}.
\ee
This line of reasoning, however, only leads to an order of magnitude
estimate of the uncontrolled remainder. In principle, any
dimensionless scalar function could be multiplying this estimate as
long as it does not change the scaling. For the case of non-spinning
holes, the metric perturbation has been computed to
$O(V^3)$~\cite{Poisson:2005pi}, which allows us to compare these terms
to Eq.~(\ref{error}). This comparison suggests that the multiplicative
scalar function is roughly unity and, thus, unnecessary so that
Eq.~(\ref{error}) is indeed a good approximation to the scaling of the
uncontrolled remainders.  Clearly, this estimate is not the exact
error in the approximation, which can formally only be determined if
we know the exact functional form of the next order term. This
estimate, however, is a physically well-motivated approximation for
the uncontrolled remainders.

The outer radius of the region of validity of this approximation can
be approximately determined by studying the behavior of these error
bars.  From Fig.~\ref{metric} we see that the error bars become
considerable large (approximately $50 \%$ as big as the perturbation
itself) roughly at $r = 13 M$.  However, for $r < 4 M$ the estimated
error bars are less than $10 \%$ relative to the perturbation. This
seems to indicate that, even at $b = 10 M$ where the slow-motion
approximation begins to become inaccurate, the approximate solution is
still valid sufficiently close to the outer horizon. This estimation
of the fractional errors is by no means a formal proof of the
existence or size of the region of validity of the approximation.
However, this estimation does provide a strong argument that the
approximation is indeed valid sufficiently close to the outer horizon.
If $10 \%$ fractional error is tolerable, which would correspond to
neglecting terms of $O(V^3)$, then the outer radius of the shell is
approximately given by $r_{outer} \approx 4 M$. Clearly, as $b$ is
increased, the slow-motion approximation will become more accurate
and, thus, the perturbation will be valid inside a bigger shell with
larger outer radii.

The results of this paper are clearly valid close to the outer horizon
of the background hole, thus allowing the study of physical processes
of interest to the relativity community. For instance, we can use the
perturbation presented here to construct initial data for a binary
system near either hole. In this case, we are not interested in the
behavior of the perturbation inside the inner horizon because that
region can be excised and does not belong to the computational domain.
Furthermore, the calculations of angular momentum and mass flux across
the horizon can still be performed because they only depend on the
behavior of the perturbation near or at the outer horizon.

\section{Conclusions}
\label{conclusions}
We have computed a tidally perturbed metric for a spinning black hole
in the slow-motion approximation. This approximation allows us to
parameterize the NP scalar $\psi_0$ [Eq.~(\ref{psi0})] in terms of the
electric and magnetic tidal tensors of the external universe. With
this scalar we can then construct a potential $\Psi$
[Eq.~(\ref{potential})], by applying certain differential operators to
it. From this potential, we can then apply the Chrzanowski procedure
to construct a metric perturbation
[Eqs.~(\ref{final-potential}),(\ref{spin-coeff-decomp}),(\ref{diff-op-on-spin-coeff}),(\ref{decomp-abc}),(\ref{diff-ops-on-psi-decomped2}),(\ref{MP1decomp}),(\ref{MP1})].

The metric is naturally computed in the ingoing radiation gauge and in
Kerr coordinates, which are suitable to study perturbation near the
horizon due to its horizon penetrating properties. This metric is
given explicitly in terms of scalar functions of the coordinates and
is parametrized by the mass of the background hole, its Kerr spin
parameter and electric and magnetic tidal tensors. The mass and the
spin parameters of the background hole do not have to be small
relative to each other, so in principle the metric presented here is
capable of representing tidally perturbed extremal Kerr holes. The
tidal tensors describe the time-evolution of the perturbation produced
by the external universe and, thus, are functions of time that should
be determined by matching the metric to another approximation valid
far from the holes.

The slow motion approximation constrains what kind of perturbations
are allowed, thus limiting the external universe that is producing
them. In this approximation, the radius of curvature of the external
universe must be changing sufficiently slowly relative to the scales
of the background hole. One consequence of this restriction is that
the tidal fields must be slowly-varying functions of time, thus
allowing us to neglect their time derivatives. In this sense, this
approximation leads to a quasi-static limit, where in this paper we
have calculated the first non-vanishing deviations from staticity.
Another consequence of this approximation is that we can parametrized
the perturbation in terms of multipole moments, where here we have only
considered the first non-vanishing one (the quadrupolar perturbation).
In perturbation theory, the $l+1$ mode will be one order smaller than
the $l$ mode, which allows us to neglect the octopole and any other
higher modes, as well as any mode beating between the quadrupole and
higher modes.

Due to these restrictions, if we allow the external universe to be
given by another hole in a quasicircular orbit around the background
(a binary system), we are limited to those whose separation is
sufficiently large. In this case, we must have sufficiently large
orbital separations so that the Riemann curvature produced by the
companion is large relative to the scales of the background hole. In
other words, this approximation will break if we consider systems that
are close to their innermost stable orbit and ready to plunge. We have
seen, however, that for a separation of $b = 10 M$, the metric
presented here is valid inside a shell given approximately by $0.01 M
< r < 4 M$.

The reason why the region of validity of the approximation is a shell
can be traced back to the choice of tetrad and to the limitations of
the Chrzanowski procedure. The perturbation cannot be valid too close
to the background hole because in that region the Kinnersly tetrad,
from which the perturbation was constructed, is also divergent. This
perturbation is also divergent as $r \to \infty$ because the
Chrzanowski procedure builds the perturbation as a linear expansion of
the metric. In other words, we cannot analyze the dynamics of the
entire spacetime with this metric, since its validity is limited to
field points sufficiently close to the outer horizon of the background
hole. However, if this metric were to be asymptotically matched to
another metric valid far from the background hole, then the resultant
metric would be accurate on the entire $3$-manifold up to uncontrolled
remainders thus, capable of reproducing the dynamics of the entire
spacetime.

This metric might be useful to different areas in general relativity.
On the one hand it might be useful in the construction of
astrophysically realistic initial data for binary systems of spinning
black holes. Its importance relies in that it could accurately
represent the metric field in the neighborhood of a black hole tidally
disrupted by a companion. These tides are analytic and arise as a true
approximate solution to the Einstein equations in the slow-motion
approximation. Initial data, however, requires a metric that is valid
in the entire $3$-manifold. Therefore, in order to use these results
as initial data they will first have to be asymptotically matched
\cite{Alvi:1999cw,Yunes:2005nn} to a PN expansion valid far from the
background holes \cite{Blanchet:2002av}. In this manner, initial data
could be constructed that satisfies the full set of the Einstein
equation, including the constraints, to a high order of accuracy.

Another use for the metric computed in this paper relates to the flux
of mass and angular momentum through a perturbed Kerr horizon. This
flux will be important for EMRIs, where the effect of tidal
perturbations could be large enough to lead to large fluxes, which in
turn could affect the gravitational wave signal emitted by the system
\cite{Martel:2003jj,Hughes:2001jr}. Recent investigations
\cite{Poisson:2004cw} have used a curvature formalism to compute this
flux directly from $\psi_0$.  However, there exists a metric formalism
to obtain this flux directly from the metric itself. An interesting
research direction would be to compute this flux and compare to the
results obtained with the curvature formalism.

Finally, the perturbed metric computed here can also be of use to the
data analysis community to construct gravitational waveforms. EMRIs
are particularly good candidates to be observed by LISA, but such
observations require extremely accurate formulae for the phasing of
the gravitational waves due to the use of matched filtering. Recently,
Ref.~\cite{Glampedakis:2005cf} studied how to use and implement a
quasi-Kerr metric (a perturbed Kerr metric in the limit of slow
rotation of the background hole) to detect EMRIs with LISA. A similar
study could be performed with the perturbed metric computed in this
paper, which can also describe rapidly rotating black holes.

Future work will concentrate on performing the necessary asymptotic
matching to shape this metric into useful initial data for numerical
relativity applications. The matching procedure will provide
expressions for the tidal tensors in terms of PN quantities, as well
as a coordinate transformation between Kerr coordinates and the
coordinate system used in the PN approximation. In this manner, a
piece-wise global solution can be computed, which will contain small
discontinuities inside the matching region that could be eliminated by
the introduction of transition function. Since these discontinuities
will be small due to the matching, the transition functions will not
alter the content of the data to the order of the approximation used.
After the matching is completed, we will have obtained an approximate
analytic global metric that will contain the tidal fields of one hole
on the other near the outer horizon of the former, where these fields
come directly from solutions to the Einstein equations.

\begin{acknowledgments}
  
  We are indebted to Eric Poisson, whose good suggestions and clear
  explanations contributed greatly to this work. We are also grateful
  to Bernd Br\"{u}gmann, Amos Ori, Ben Owen and Gerhard Sch\"{a}fer
  for useful discussions and comments. Finally, we thank Bernd
  Br\"{u}gmann and Ben Owen for their continuous support and
  encouragement. N.Y would also like to thank the University of Jena
  for their hospitality.
  
  This work was supported by the Institute for Gravitational Physics
  and Geometry and the Center for Gravitational Wave Physics, funded
  by the National Science Foundation under Cooperative Agreement
  PHY-01-14375. This work was also supported by NSF grants
  PHY-02-18750, PHY-02-44788, and PHY-02-45649, as well as the DFG
  grant ``SFB Transregio 7: Gravitationswellenastronomie.''

\end{acknowledgments}

\section*{Appendix}
\label{appendix}
In this appendix we provide explicit formula for the transformation
from Kerr coordinates to Kerr-Schild coordinate. This transformation
can be found in Refs.~\cite{MTW,Carrol,Dinverno} and is given by
\ba
x &=& \sin{\theta} \left( r \cos{\phi} - a \sin{\phi} \right),
\nonumber \\
y &=& \sin{\theta} \left( r \sin{\phi} + a \cos{\phi} \right).
\nonumber \\
z &=& r \cos{\theta}. 
\ea
The inverse transformation is given by 
\ba
r &=& \sqrt{\frac{R^2 - a^2 + w}{2}},
\nonumber \\
w &=& \sqrt{(R^2 - a^2)^2 + 4 a^2 z^2},
\nonumber \\
\theta &=& \arccos{\frac{z}{r}},
\nonumber \\
\phi &=& \arctan{\frac{r y - a x}{r x + a y}}.
\ea
Other useful relations are
\ba
\sin{\theta} &=& \left(\frac{x^2 + y^2}{r^2 + a^2} \right)^{1/2},
\nonumber \\
\sin{\phi} &=& \frac{ r y - a x}{\left[ (r^2 + a^2) (x^2 + y^2) \right]^{1/2}},
\nonumber \\
\cos{\phi} &=&  \frac{ r x + a y}{\left[ (r^2 + a^2) (x^2 + y^2) \right]^{1/2}}.
\ea
Note that these transformation reduce to the usual transformation from
spherical polar coordinates to Cartesian coordinates in the limit $a
\to 0$. 

The Jacobian of the transformation, $\Lambda^a{}_b = \partial
x^a/\partial x^b$, is given explicitly by 
\ba
\Lambda^r{}_x &=& \frac{x}{2 r} \left( 1 + \frac{R^2 - a^2}{w} 
\right),
\nonumber \\
\Lambda^r{}_y &=& \frac{y}{2 r} \left( 1 + \frac{R^2 - a^2}{w}
\right) , 
\nonumber \\
\Lambda^r{}_z &=& \frac{z}{2 r} \left( 1 + \frac{R^2 + a^2}{w}
\right),
\nonumber \\
\Lambda^{\theta}{}_x &=& \frac{z x}{2 r^2} \left(1 + \frac{R^2 -
    a^2}{w} \right) \left( r^2 - z^2 \right)^{-1/2},
\nonumber \\
\Lambda^{\theta}{}_y &=& \frac{z y}{2 r^2} \left(1 + \frac{R^2 -
    a^2}{w} \right) \left( r^2 - z^2 \right)^{-1/2},
\nonumber \\
\Lambda^{\theta}{}_z &=& - \left( r^2 - z^2 \right)^{-1/2} \left[ 1
  - \frac{z^2}{2 r^2} \left( 1 + \frac{R^2 + a^2}{w} \right) \right],
\nonumber \\
\Lambda^{\phi}{}_x &=& \frac{r x + a y}{(r^2 + a^2)(x^2 + y^2)}
\left[ \left( y - x \frac{r y - a x}{r x + a y} \right) \left( 1 
\right. \right. 
\nonumber \\
&& \left. \left.
+  \frac{R^2 - a^2}{w} \right) \frac{x}{2r} - a - r \frac{r y - a
    x}{rx + ay} \right],
\nonumber \\
\Lambda^{\phi}{}_y &=& \frac{r x + a y}{(r^2 + a^2)(x^2 + y^2)}
\left[ \left( y - x \frac{r y - a x}{r x + a y} \right) \left( 1 
\right. \right.
\nonumber \\
&& \left. \left.
+  \frac{R^2 - a^2}{w} \right) \frac{y}{2r} + r - a \frac{r y - a
    x}{rx + ay} \right],
\nonumber \\
\Lambda^{\phi}{}_z &=& \frac{r x + a y}{(r^2 + a^2)(x^2 + y^2)}
\left[ \left( y - x \frac{r y - a x}{r x + a y} \right) \left( 1 
\right. \right. 
\nonumber \\
&& \left. \left. 
+  \frac{R^2 + a^2}{w} \right) \frac{z}{2r} \right].
\nonumber \\
\ea
Note that this Jacobian reduces to the standard Jacobian of the
transformation between spherical polar and Cartesian coordinates in
the limit $a \to 0$. There is a more elegant way to write this
Jacobian in tensor notation as
\ba
\Lambda^r{}_a &=& \frac{1}{2r} \left\{ \delta_{ai} x^i + \frac{1}{w}
  \left[ \delta_{ia} x^i \left( R^2 - a^2 \right) + 2 a^2 z
    \delta_{az} \right] \right\},
\nonumber \\
\Lambda^{\theta}{}_a &=& \frac{z}{r \left(r^2 - a^2\right)^{1/2}}
\Lambda^r{}_a,
\nonumber \\
\Lambda^{\phi}{}_a &=& \frac{r x + a y}{(r^2  + a^2) (x^2 + y^2)}
\left[ \Lambda^r{}_a  \left( y - \frac{r y - a x}{r x + a y} x \right) 
\right.
\\ \nonumber 
&& \left. 
+  \left( r - \frac{r y - a x}{r x + a y} a \right) \delta_{ya} -
  \left( a + \frac{ ry - a x}{r x + a y} r \right) \delta_{xa} \right].
\ea

With this Jacobian, the metric in Kerr-Schild coordinates is given by
\be
g_{ab} = g_{a'b'} \Lambda^{a'}{}_a \Lambda^{b'}{}_b,
\ee
where here the primed indices refer to spherical coordinates and the
unprimed indices to Cartesian coordinates. 

\bibliography{paper.bib}

\end{document}